\documentclass[prb,letterpaper,twocolumn]{revtex4}
\usepackage{epsfig}
\usepackage{times}

\begin{document}
\title{ APPLICATIONS OF PHYSICS TO ECONOMICS AND FINANCE:\\ MONEY,
 INCOME, WEALTH, AND THE STOCK MARKET \footnote{This document is a
 reformatted version of my PhD thesis.  Advisor: Professor Victor M.\
 Yakovenko. Committee: Professor J.\ Robert Dorfman, Professor
 Theodore L.\ Einstein, Professor Bei-Lok Hu, and Professor John D.\
 Weeks.}}

\author{Adrian A.\ Dr\u{a}gulescu}
\email[Email:]{adrian.dragulescu@constellation.com}
\affiliation{Department of Physics, University of Maryland, College
  Park }
\thanks{}

\date{May 15, 2002, cond-mat/0307341}

\begin{abstract}
  {\bf Abstract:} Several problems arising in Economics and Finance
  are analyzed using concepts and quantitative methods from Physics.
  The disertation is organized as follows:

  In the first chapter, it is argued that in a closed economic system,
  money is conserved.  Thus, by analogy with energy, the equilibrium
  probability distribution of money must follow the exponential
  Boltzmann-Gibbs law characterized by an effective temperature equal
  to the average amount of money per economic agent.  The emergence of
  Boltzmann-Gibbs distribution is demonstrated through computer
  simulations of economic models.  A thermal machine which extracts a
  monetary profit can be constructed between two economic systems with
  different temperatures.  The role of debt and models with broken
  time-reversal symmetry for which the Boltzmann-Gibbs law does not
  hold, are discussed.

  In the second chapter, using data from several sources, it is found
  that the distribution of income is described for the great majority
  of population by an exponential distribution, whereas the high-end
  tail follows a power law.  From the individual income distribution,
  the probability distribution of income for families with two earners
  is derived and it is shown that it also agrees well with the data.
  Data on wealth is presentend and it is found that the distribution
  of wealth has a structure similar to the distribution of income.
  The Lorenz curve and Gini coefficient were calculated and are shown
  to be in good agreement with both income and wealth data sets.

  In the third chapter, the stock-market fluctuations at different
  time scales are investigated.  A model where stock-price dynamics is
  governed by a geometrical (multiplicative) Brownian motion with
  stochastic variance is proposed.  The corresponding Fokker-Planck
  equation can be solved exactly. Integrating out the variance, an
  analytic formula for the time-dependent probability distribution of
  stock price changes (returns) is found.  The formula is in excellent
  agreement with the Dow-Jones index for the time lags from 1 to 250
  trading days.  For time lags longer than the relaxation time of
  variance, the probability distribution can be expressed in a scaling
  form using a Bessel function.  The Dow-Jones data follow the scaling
  function for seven orders of magnitude.
\end{abstract}
\maketitle

\tableofcontents

\section*{Acknowledgements}
My six year apprenticeship in physics at the University of Maryland
was a unique and remarkable leg of my life.  And to a large degree
this is due to my advisor, Professor Victor M. Yakovenko, who has
been at the center of my PhD education.  I will deeply miss his
clear perspective, his sure hand, and the incommensurate thrill of
doing physics together.  

I am grateful to Professors J.\ Robert Dorfman, Theodore L.\
Einstein, Bei-Lok Hu, and John D.\ Weeks for agreeing to serve on 
my advisory committee. 

Thanks are also due to Krishnendu Sengupta, Nicholas Dupuis, Hyok-Jon
Kwon, Anatoley Zheleznyak, and Hsi-Sheng Goan all former students or
post-docs of professor Victor Yakovenko, for stimulating discussions
in condensed matter physics.  My interests in physics were greatly
molded by professor Bei-Lok Hu, through his remarkable courses in
non-equilibrium physics, and also through personal interaction for
more than three years.  I had profited a lot from the association with
Charis Anastopoulos and Sanjiv Shresta while working on quantum optics 
problems suggested by professor Bei-Lok Hu.

It is not without melancholy that I thank my first true physics
teachers: Ion Cot\u{a}escu, Ovidiu Lipan, Dan Luca, Adrian Neagu,
Mircea Ra\c{s}a, Costel Ra\c{s}inariu, Vlad Socoliuc, and Dumitru
Vulcanov.  On an even deeper layer, I thank my family for all the
support they provided me over the years, and for allowing me to pursue
my interests.

All the people mentioned above have generously shared with me some of
their time and knowledge.  I will do all I can to represent them well.

\section{Statistical mechanics of money}  
\label{ch:1}
\subsection{Introduction}

The application of statistical physics methods to econo\-mics promises
fresh insights into problems traditionally not associated with physics
(see, for example, the recent review and book \cite{Farmer}).  Both
statistical mechanics and economics study big ensembles: collections
of atoms or economic agents, respectively.  The fundamental law of
equilibrium statistical mechanics is the Boltzmann-Gibbs law, which
states that the probability distribution of energy $\varepsilon$ is
$P(\varepsilon)=Ce^{-\varepsilon/T}$, where $T$ is the temperature,
and $C$ is a normalizing constant \cite{StatPhys}.  The main
ingredient that is essential for the textbook derivation of the
Boltzmann-Gibbs law \cite{StatPhys} is the conservation of energy
\cite{Tsallis}.  Thus, one may generalize that any conserved quantity
in a big statistical system should have an exponential probability
distribution in equilibrium.

An example of such an unconventional Boltzmann-Gibbs law is the
probability distribution of forces experienced by the beads in a
cylinder pressed with an external force \cite{Nagel}.  Because the
system is at rest, the total force along the cylinder axis experienced
by each layer of granules is constant and is randomly distributed
among the individual beads.  Thus the conditions are satisfied for the
applicability of the Boltzmann-Gibbs law to the force, rather than
energy, and it was indeed found experimentally \cite{Nagel}.

We claim that, in a closed economic system, the total amount of money
is conserved.  Thus the equilibrium probability distribution of money
$P(m)$ should follow the Boltzmann-Gibbs law $P(m)=Ce^{-m/T}$.  Here
$m$ is money, and $T$ is an effective temperature equal to the average
amount of money per economic agent.  The conservation law of money
\cite{Shubik} reflects their fundamental property that, unlike
material wealth, money (more precisely the fiat, ``paper'' money) is
not allowed to be manufactured by regular economic agents, but can
only be transferred between agents.  Our approach here is very similar
to that of Ispolatov {\it et al.}  \cite{Redner}.  However, they
considered only models with broken time-reversal symmetry, for which
the Boltzmann-Gibbs law typically does not hold.  The role of
time-reversal symmetry and deviations from the Boltzmann-Gibbs law are
discussed in detail in Sec.\ \ref{Non-Gibbs}.

It is tempting to identify the money distribution $P(m)$ with the
distribution of wealth \cite{Redner}.  However, money is only one part
of wealth, the other part being material wealth.  Material products
have no conservation law because they can be manufactured, destroyed, 
consumed, etc.  Moreover, the monetary value of a material product
(the price) is not constant.  The same applies to stocks, which
economics textbooks explicitly exclude from the definition of money
\cite{McConnell}.  So, we do not expect the 
Boltzmann-Gibbs law for the distribution of wealth.  Some authors
believe that wealth is distributed according to a power law
(Pareto-Zipf), which originates from a multiplicative random process
\cite{Montroll1}.  Such a process may reflect, among other things, the
fluctuations of prices needed to evaluate the monetary value of
material wealth.

\subsection{Boltzmann-Gibbs distribution}

Let us consider a system of many economic agents $N\gg1$, which may be
individuals or corporations.  In this thesis, we only consider the case
where their number is constant.  Each agent $i$ has some money $m_i$
and may exchange it with other agents.  It is implied that money is
used for some economic activity, such as buying or selling material
products; however, we are not interested in that aspect.  As in Ref.\
\cite{Redner}, for us the only result of interaction between agents
$i$ and $j$ is that some money $\Delta m$ changes hands:
$[m_i,m_j]\to[m_i',m_j']=[m_i-\Delta m,m_j+\Delta m]$.  Notice that
the total amount of money is conserved in each transaction:
$m_i+m_j=m_i'+m_j'$.  This local conservation law of money
\cite{Shubik} is analogous to the conservation of energy in collisions
between atoms.  We assume that the economic system is closed, i.\ e.\
there is no external flux of money, thus the total amount of money $M$
in the system is conserved.  Also, in the first part of this chapter,
we do not permit any debt, so each agent's money must be non-negative:
$m_i\geq0$.  A similar condition applies to the kinetic energy of
atoms: $\varepsilon_i\geq0$.

Let us introduce the probability distribution function of money
$P(m)$, which is defined so that the number of agents with money
between $m$ and $m+dm$ is equal to $NP(m)\,dm$.  We are interested in
the stationary distribution $P(m)$ corresponding to the state of
thermodynamic equilibrium.  In this state, an individual agent's money
$m_i$ strongly fluctuates, but the overall probability distribution
$P(m)$ does not change.

The equilibrium distribution function $P(m)$ can be derived in the
same manner as the equilibrium distribution function of energy
$P(\varepsilon)$ in physics \cite{StatPhys}.  Let us divide the system
into two subsystems 1 and 2.  Taking into account that money is
conserved and additive: $m=m_1+m_2$, whereas the probability is
multiplicative: $P=P_1P_2$, we conclude that
$P(m_1+m_2)=P(m_1)P(m_2)$.  The solution of this equation is
$P(m)=Ce^{-m/T}$; thus the equilibrium probability distribution of
money has the Boltzmann-Gibbs form.  From the normalization conditions
$\int_0^\infty P(m)\,dm=1$ and $\int_0^\infty m\,P(m)\,dm=M/N$, we
find that $C=1/T$ and $T=M/N$.  Thus, the effective temperature $T$ is
the average amount of money per agent.

The Boltzmann-Gibbs distribution can be also derived by maximizing
\hfill the \hfill entropy \hfill of \hfill money \hfill distribution \\ 
$S=-\int_0^\infty dm\,P(m)\ln P(m)$
under the constraint of mon\-ey con\-ser\-vation \cite{StatPhys}.  Following
original Boltzmann's argument, let us divide the money axis $0\leq
m\leq\infty$ into small bins of size $dm$ and number the bins
consecutively with the index $b=1,2,\dots\;$ Let us denote the number
of agents in a bin $b$ as $N_b$, the total number being
$N=\sum_{b=1}^{\infty}N_b$.  The agents in the bin $b$ have money
$m_b$, and the total money is $M=\sum_{b=1}^{\infty}m_bN_b$.  The
probability of realization of a certain set of occupation numbers
$\{N_b\}$ is proportional to the number of ways $N$ agents can be
distributed among the bins preserving the set $\{N_b\}$.  This number
is $N!/N_1!N_2!\ldots\;$ The logarithm of probability is entropy $\ln
N!-\sum_{b=1}^{\infty}\ln N_b!$.  When the numbers $N_b$ are big and
Stirling's formula $\ln N!\approx N\ln N - N$ applies, the entropy per
agent is $S=(N\ln N-\sum_{b=1}^{\infty}N_b\ln
N_b)/N=-\sum_{b=1}^{\infty}P_b\ln P_b$, where $P_b=N_b/N$ is the
probability that an agent has money $m_b$.  Using the method of
Lagrange multipliers to maximize the entropy $S$ with respect to the
occupation numbers $\{N_b\}$ with the constraints on the total money
$M$ and the total number of agents $N$ generates the Boltzmann-Gibbs
distribution for $P(m)$ \cite{StatPhys}.

\subsection{Computer simulations}

\begin{figure}
\centerline{
\epsfig{file=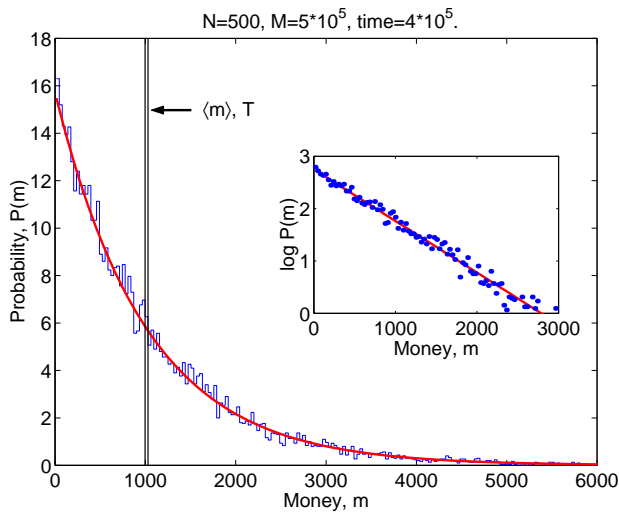,width=0.95\linewidth}}
\caption{{\small Histogram and points: stationary probability 
dis\-tri\-bu\-tion of mo\-ney $P(m)$. Solid curves: fits to the
Boltzmann-Gibbs law $P(m)\propto\exp(-m/T)$.  Vertical lines: the
initial distribution of money.}}
\label{fig:model3}
\end{figure}

To check that these general arguments indeed work, we performed
several computer simulations.  Initially, all a\-gents are given the
same amount of money: $P(m)=\delta(m-M/N)$, which is shown in Fig.\
\ref{fig:model3} as the double vertical line.  One pair of agents at a
time is chosen randomly, then one of the agents is randomly picked to
be the ``winner'' (the other agent becomes the ``loser''), and the
amount $\Delta m\geq0$ is transferred from the loser to the winner.
If the loser does not have enough money to pay ($m_i<\Delta m$), then
the transaction does not take place, and we proceed to another pair of
agents.  Thus, the agents are not permitted to have negative money.
This boundary condition is crucial in establishing the stationary
distribution.  As the agents exchange money, the initial
delta-function distribution first spreads symmetrically.  Then, the
probability density starts to pile up at the impenetrable boundary
$m=0$.  The distribution becomes asymmetric (skewed) and ultimately
reaches the stationary exponential shape shown in Fig.\
\ref{fig:model3}.  We used several trading rules in the simulations:
the exchange of a small constant amount $\Delta m=1$, the exchange of
a random fraction $0\le\nu\le1$ of the average money of the pair:
$\Delta m=\nu(m_i+m_j)/2$, and the exchange of a random fraction $\nu$
of the average money in the system: $\Delta m=\nu\,M/N$.  Figures in
this chapter mostly show simulations for the third rule; however, the
final stationary distribution was found to be the same for all rules.

\begin{figure}
\centerline{
\epsfig{file=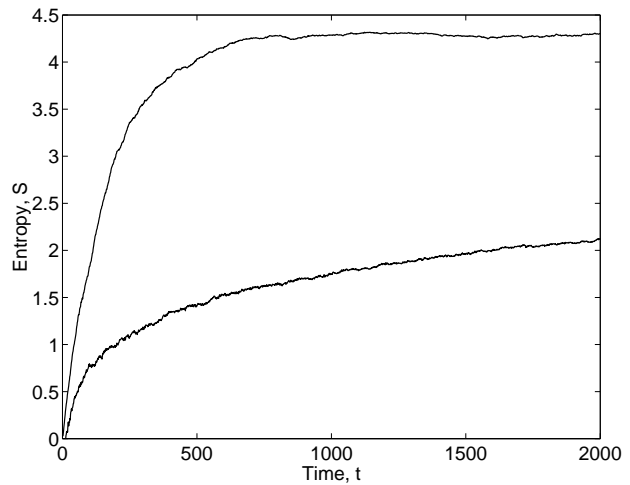,width=0.95\linewidth}}
\caption{{\small Time evolution of entropy.  Top curve: for the exchange of a
random fraction $\nu$ of the average money in the system: $\Delta
m=\nu\,M/N$.  Bottom curve: for the exchange of a small constant
amount $\Delta m=1$.  The time scale for the bottom curve is 500 times
greater than indicated, so it actually ends at the time $10^6$.}}
\label{fig:entropy}
\end{figure}

In the process of evolution, the entropy $S$ of the system increases
in time and saturates at the maximal value for the Boltzmann\-Gibbs
distribution.  This is illustrated by the top curve in Fig.\
\ref{fig:entropy} computed for the third rule of exchange.  The bottom
curve in Fig.\ \ref{fig:entropy} shows the time evolution of entropy
for the first rule of exchange.  The time scale for this curve is 500
times greater than for the top curve, so the bottom curve actually
ends at the time $10^6$.  The plot shows that, for the first rule of
exchange, mixing is much slower than for the third one.  Nevertheless,
even for the first rule, the system also eventually reaches the
Boltzmann-Gibbs state of maximal entropy, albeit over a time much
longer than shown in Fig.\ \ref{fig:entropy}.

One might argue that the pairwise exchange of money may correspond to
a medieval market, but not to a modern economy.  In order to make the
model somewhat more realistic, we introduce firms.  One agent at a
time becomes a ``firm''.  The firm borrows capital $K$ from another
agent and returns it with an interest $rK$, hires $L$ agents and pays
them wages $W$, manufactures $Q$ items of a product and sells it to $Q$
agents at a price $R$.  All of these agents are randomly selected.
The firm receives the profit $F=RQ-LW-rK$.  The net result is a
many-body exchange of money that still satisfies the conservation law.

Parameters of the model are selected following the procedure described
in economics textbooks \cite{McConnell}.  The aggregate demand-supply
curve for the product is taken to be $R(Q)=V/Q^\eta$, where $Q$ is the
quantity people would buy at a price $R$, and $\eta=0.5$ and $V=100$
are constants.  The production function of the firm has the
conventional Cobb-Douglas form: $Q(L,K)=L^\beta K^{1-\beta}$, where
$\beta=0.8$ is a constant.  In our simulation, we set $W=10$.  By
maximizing firm's profit $F$ with respect to $K$ and $L$, we find the
values of the other parameters: $L=20$, $Q=10$, $R=32$, and $F=68$.

However, the actual values of the parameters do not matter.  Our
computer simulations show that the stationary probability distribution
of money in this model always has the universal Boltzmann-Gibbs form
independent of the model parameters.

\subsection{Thermal machine}

As explained in the Introduction, the money distribution $P(m)$ should not
be confused with the distribution of wealth.  We believe that $P(m)$
should be interpreted as the instantaneous distribution of purchasing
power in the system.  Indeed, to make a purchase, one needs money.
Material wealth normally is not used directly for a purchase.  It
needs to be sold first to be converted into money.

Let us consider an outside monopolistic vendor selling a product (say,
cars) to the system of agents at a price $p$.  Suppose that a certain
small fraction $f$ of the agents needs to buy the product at a given
time, and each agent who has enough money to afford the price will buy
one item.  The fraction $f$ is assumed to be sufficiently small, so
that the purchase does not perturb the whole system significantly.  At
the same time, the absolute number of agents in this group is assumed
to be big enough to make the group statistically representative and
characterized by the Boltzmann-Gibbs distribution of money.  The
agents in this group continue to exchange money with the rest of the
system, which acts as a thermal bath.  The demand for the product is
constantly renewed, because products purchased in the past expire
after a certain time.  In this situation, the vendor can sell the
product persistently, thus creating a small steady leakage of money
from the system to the vendor.

What price $p$ would maximize the vendor's income?  To answer this
question, it is convenient to introduce the cumulative distribution of
purchasing power ${\cal N}(m)=N\int_m^\infty P(m')\,dm'=Ne^{-m/T}$,
which gives the number of agents whose money is greater than $m$.  The
vendor's income is $fp{\cal N}(p)$.  It is maximal when $p=T$, i.\ e.\
the optimal price is equal to the temperature of the system.  This
conclusion also follows from the simple dimensional argument that
temperature is the only money scale in the problem.  At the price
$p=T$ that maximizes the vendor's income, only the fraction ${\cal
N}(T)/N=e^{-1}=0.37$ of the agents can afford to buy the product.

Now let us consider two disconnected economic systems, one with the
temperature $T_1$ and another with $T_2$: $T_1>T_2$.  A vendor can buy
a product in the latter system at its equilibrium price $T_2$ and sell
it in the former system at the price $T_1$, thus extracting the
speculative profit $T_1-T_2$, as in a thermal machine.  This example
suggests that speculative profit is possible only when the system as a
whole is out of equilibrium.  As money is transferred from the high-
to the low-temperature system, their temperatures become closer and
eventually equal.  After that, no speculative profit is possible,
which would correspond to the ``thermal death'' of the economy.  This
example brings to mind economic relations between developed and
developing countries, with manufacturing in the poor (low-temperature)
countries for export to the rich (high-temperature) ones.

We will demonstrate in Ch.\ \ref{ch:2} that for the large majority of
the population the distribution of income is exponential.  Hence,
similar to the distribution of money, the distribution of income has a
corresponding temperature.  If in the previous discussion about
economic trade, the two countries have a different temperature for the
distribution of income, the possibility of constructing a thermal
machine will still hold true.  This is because purchasing power (total
money) per unit time has two components: a positive one, income and a
negative one, spending.  Instead of making a one time purchase, the
buyer will buy one product per unit time, effectively creating the
thermal engine.  We will revisit this idea in Sec.\ \ref{geo} when we
give the values for income temperature between the states of the USA
and for the United Kingdom.

\subsection{Models with debt}

\begin{figure}
\centerline{
\epsfig{file=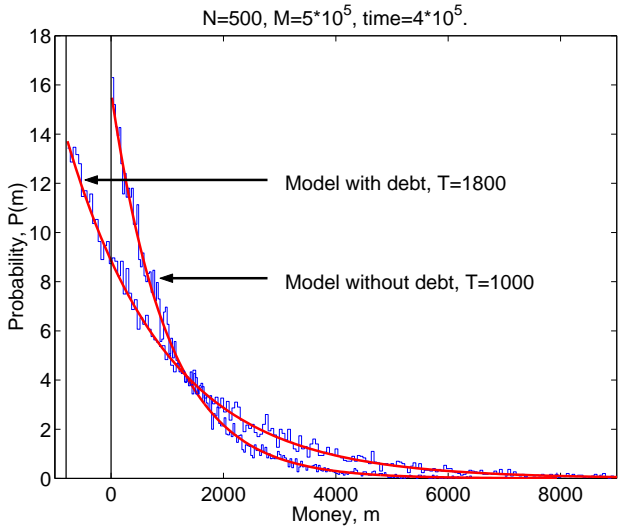,width=0.95\linewidth}}
\caption{{\small Histograms: stationary distributions of money with and
without debt.  Solid curves: fits to the Boltzmann-Gibbs laws with
temperatures $T=1800$ and $T=1000$.}}
\label{fig:modeldebt}
\end{figure}

Now let us discuss what happens if the agents are permitted to go into
debt.  Debt can be viewed as negative money.  Now when a loser does
not have enough money to pay, he can borrow the required amount from a
reservoir, and his balance becomes negative.  The conservation law is
not violated: The sum of the winner's positive money and loser's
negative money remains constant.  When an agent with a negative
balance receives money as a winner, she uses this money to repay the
debt until her balance becomes positive.  We assume for simplicity
that the reservoir charges no interest for the lent money.  However,
because it is not sensible to permit unlimited debt, we put a limit
$m_d$ on the maximal debt of an agent: $ m_i>-m_d$.  This new boundary
condition $P(m<-m_d)=0$ replaces the old boundary condition
$P(m<0)=0$.  The result of a computer simulation with $m_d=800$ is
shown in Fig.\ \ref{fig:modeldebt} together with the curve for
$m_d=0$.  $P(m)$ is again given by the Boltzmann-Gibbs law, but now
with the higher temperature $T=M/N+m_d$, because the normalization
conditions need to be maintained including the population with
negative money: $\int_{-m_d}^\infty P(m)\,dm=1$ and
$\int_{-m_d}^\infty m\,P(m)\,dm=M/N$.  The higher temperature makes
the money distribution broader, which means that debt increases
inequality between agents.

In general, temperature is completely determined by the average money
per agent, $\langle m\rangle=M/N$, and the boundary conditions.
Suppose the agents are required to have no less money than $m_1$ and
no more than $m_2$: $m_1\leq m\leq m_2$.  In this case, the
two normalization conditions: $\int_{m_1}^{m_2}P(m)\,dm=1$ and
$\int_{m_1}^{m_2} m\,P(m)\,dm=\langle m\rangle$ with
$P(m)=C\,e^{-m/T}$ give the following equation for $T$ 
\begin{equation}
  \coth\left(\frac{\Delta m}{T}\right)-\frac{T}{\Delta m}=
  \frac{\overline{m}-\langle m\rangle}{\Delta m},
\label{T}
\end{equation}
where $\overline{m}=(m_1+m_2)/2$ and $\Delta m=(m_2-m_1)/2$.  It
follows from Eq.\ (\ref{T}) that the temperature is positive when
$\overline{m}>\langle m\rangle$, negative when $\overline{m}<\langle
m\rangle$, and infinite ($P(m)=\rm const$) when $\overline{m}=\langle
m\rangle$.  In particular, if agents' money are bounded from above,
but not from below: $-\infty\leq m\leq m_2$, the temperature is
negative.  That means an inverted Boltzmann-Gibbs distribution with
more rich agents than poor.

Imposing a sharp cutoff at $m_d$ may be not quite realistic.  In
practice, the cutoff may be extended over some range depending on the
exact bankruptcy rules.  Over this range, the Boltzmann-Gibbs
distribution would be smeared out.  So we expect to see the
Boltzmann-Gibbs law only sufficiently far from the cutoff region.
Similarly, in experiment \cite{Nagel}, some deviations from the
exponential law were observed near the lower boundary of the
distribution.  Also, at the high end of the distributions, the number
of events becomes small and statistics poor, so the Boltzmann-Gibbs
law loses applicability.  Thus, we expect the Boltzmann-Gibbs law to
hold only for the intermediate range of money not too close either to
the lower boundary or to the very high end.  However, this range is
the most relevant, because it covers the great majority of population.

Lending creates equal amounts of positive (asset) and negative
(liability) money \cite{Shubik,McConnell}.  When economics textbooks
describe how ``banks create money'' or ``debt creates money''
\cite{McConnell}, they do not count the negative liabilities as money,
and thus their money is not conserved.  In our operational definition
of money, we include all financial instruments with fixed
denomination, such as currency, IOUs, and bonds, but not material
wealth or stocks, and we count both assets and liabilities.  With this
definition, money is conserved, and we expect to see the
Boltzmann-Gibbs distribution in equilibrium.  Unfortunately, because
this definition differs from economists' definitions of money (M1, M2,
M3, etc.\ \cite{McConnell}), it is not easy to find the appropriate
statistics.  Of course, money can be also emitted by a central bank or
government.  This is analogous to an external influx of energy into a
physical system.  However, if this process is sufficiently slow, the
economic system may be able to maintain quasi-equilibrium,
characterized by a slowly changing temperature.

We performed a simulation of a model with one bank and many agents.
The agents keep their money in accounts on which the bank pays
interest.  The agents may borrow money from the bank, for which they
must pay interest in monthly installments.  If they cannot make the
required payments, they may be declared bankrupt, which relieves them
from the debt, but the liability is transferred to the bank.  In this
way, the conservation of money is maintained.  The model is too
elaborate to describe it in full detail here.  We found that,
depending on the parameters of the model, either the agents constantly
lose money to the bank, which steadily reduces the agents'
temperature, or the bank constantly loses money, which drives down its
own negative balance and steadily increases the agents' temperature.

\subsection{Boltzmann equation}

The Boltzmann-Gibbs distribution can be also derived from the
Boltzmann equation \cite{Kinetics}, which describes the time evolution
of the distribution function $P(m)$ due to pairwise interactions: 
\begin{eqnarray} 
  &&\!\!\!\!\!\!\!\!\!\frac{dP(m)}{dt}=\int\!\!\int\{
    -w_{[m,m']\to[m-\Delta,m'+\Delta]}P(m)P(m')
\label{Boltzmann}  \\
  &&+w_{[m-\Delta,m'+\Delta]\to[m,m']}
  P(m-\Delta)P(m'+\Delta)\}\,dm'\,d\Delta.  \nonumber
\end{eqnarray} 
Here $w_{[m,m']\to[m-\Delta,m'+\Delta]}$ is the rate of transferring
mo\-ney $\Delta$ from an agent with money $m$ to an agent with money
$m'$.  If a model has time-reversal symmetry, then the transition rate
of a direct process is the same as the transition rate of the reversed
process, thus the $w$-factors in the first and second lines of Eq.\
(\ref{Boltzmann}) are equal.  In this case, it can be easily checked
that the Boltzmann-Gibbs distribution $P(m)=C\exp(-m/T)$ nullifies the
right-hand side of Eq.\ (\ref{Boltzmann}); thus this distribution is
stationary: $dP(m)/dt=0$
\cite{Kinetics}.

\subsection{Non-Boltzmann-Gibbs distributions}
\label{Non-Gibbs}

\begin{figure}
\centerline{
\epsfig{file=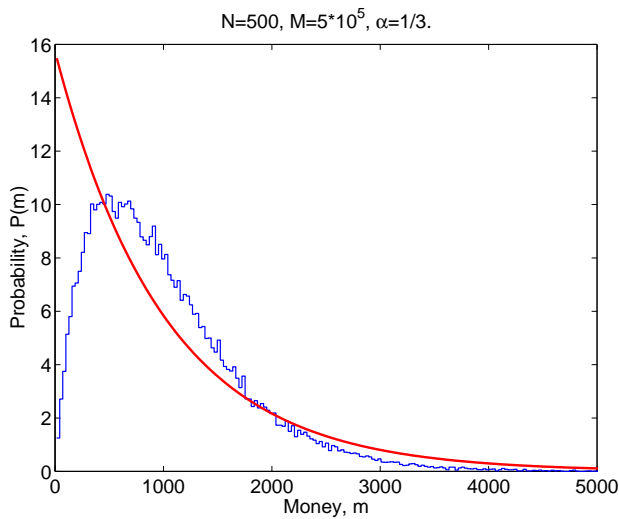,width=0.95\linewidth}}
\caption{{\small Histogram: stationary probability distribution of money in
the multiplicative random exchange model studied in Ref.\
\protect\cite{Redner}.  Solid curve: the Boltzmann-Gibbs law.}}
\label{fig:rednerplot}
\end{figure}

However, if time-reversal symmetry is broken, the two transition rates
$w$ in Eq.\ (\ref{Boltzmann}) may be different, and the system may
have a non-Boltzmann-Gibbs stationary distribution or no stationary
distribution at all.  Examples of this kind were studied in Ref.\
\cite{Redner}.  One model was called multiplicative random
exchange.  In this model, a randomly selected loser $i$ loses a fixed
fraction $\alpha$ of his money $m_i$ to a randomly selected winner
$j$: $[m_i,m_j]\to[(1-\alpha)m_i\:,\:m_j+\alpha m_i]$.  If we try to
reverse this process and immediately appoint the winner $j$ to become
a loser, the system does not return to the original configuration
$[m_i,m_j]$: $[(1-\alpha)m_i\:,\:m_j+\alpha
m_i]\to[(1-\alpha)m_i+\alpha(m_j+\alpha m_i)\:,\:(1-\alpha)(m_j+\alpha
m_i)]$.  Except for $\alpha=1/2$, the exponential distribution
function is not a stationary solution of the Boltzmann equation
derived for this model in Ref.\ \cite{Redner}.  Instead, the
stationary distribution has the shape shown in Fig.\
\ref{fig:rednerplot} for $\alpha=1/3$, which we reproduced in our
numerical simulations.  It still has an exponential tail end at the
high end, but drops to zero at the low end for $\alpha<1/2$.  Another
example of similar kind was studied in Ref.\ \cite{Chakraborti}, which
appeared after the first version of our paper was posted as
cond-mat/0001432 on January 30, 2000.  In that model, the agents save
a fraction $\lambda$ of their money and exchange a random fraction
$\epsilon$ of their total remaining money: $[m_i, m_j] \to [\lambda
m_i + \epsilon(1-\lambda)(m_i+m_j)\:,\: \lambda m_j +
(1-\epsilon)(1-\lambda)(m_i+m_j)]$.  This exchange also does not
return to the original configuration after being reversed.  The
stationary probability distribution was found in Ref.\
\cite{Chakraborti} to be nonexponential for $\lambda\neq0$ with a
shape qualitatively similar to the one shown in Fig.\
\ref{fig:rednerplot}.

\begin{figure}
\centerline{
\epsfig{file=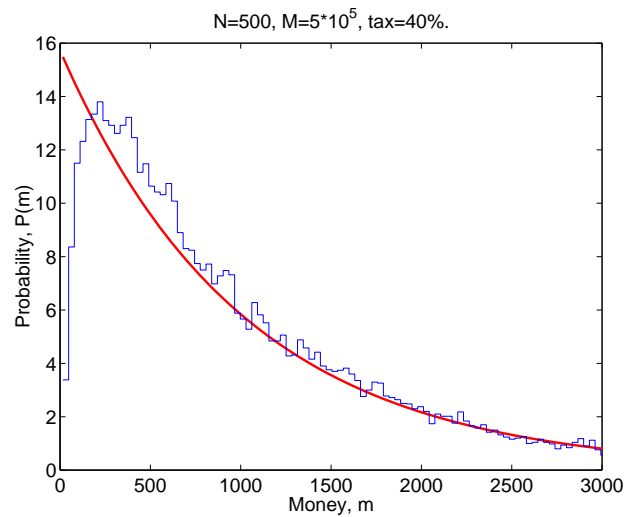,width=0.95\linewidth}}
\caption{{\small Histogram: stationary probability distribution of money in
the model with taxes and subsidies.  Solid curve: the Boltzmann-Gibbs
law.}}
\label{fig:govmod}
\end{figure}

Another interesting example which has a non-Boltz\-mann-Gibbs distribution
occurs in a model with taxes and subsidies.  Suppose a special agent
(``government'') collects a fraction (``tax'') of every transaction in
the system.  The collected money is then equally divided among all
agents of the system, so that each agent receives the subsidy $\delta
m$ with the frequency $1/\tau_s$.  Assuming that $\delta m$ is small
and approximating the collision integral with a relaxation time
$\tau_r$ \cite{Kinetics}, we obtain the following Boltzmann equation
\begin{equation} 
  \frac{\partial P(m)}{\partial t}+\frac{\delta m}{\tau_s}\,
  \frac{\partial P(m)}{\partial m}=-\frac{P(m)-\tilde{P}(m)}{\tau_r},
\label{tax}
\end{equation} 
where $\tilde{P}(m)$ is the equilibrium Boltzmann-Gibbs function.  The
second term in the left-hand side of Eq.\ (\ref{tax}) is analogous to
the force applied to electrons in a metal by an external electric
field \cite{Kinetics}.  The approximate stationary solution of Eq.\
(\ref{tax}) is the displaced Boltzmann-Gibbs one:
$P(m)=\tilde{P}(m-(\tau_r/\tau_s)\,\delta m)$.  The displacement of
the equilibrium distribution $\tilde{P}(m)$ by
$(\tau_r/\tau_s)\,\delta m$ would leave an empty gap near $m=0$.  This
gap is filled by interpolating between zero population at $m=0$ and
the displaced distribution.  The curve obtained in a computer
simulation of this model (Fig.\ \ref{fig:govmod}) qualitatively agrees
with this expectation.  The low-money population is suppressed,
because the government, acting as an external force, ``pumps out''
that population and pushes the system out of thermodynamic
equilibrium.  We found that the entropy of the stationary state in the
model with taxes and subsidies is few a percent lower than without.

These examples show that the Boltzmann-Gibbs distribution is not fully
universal, meaning that it does not hold for just any model of
exchange that conserves money.  Nevertheless, it is universal in a
limited sense: For a broad class of models that have time-reversal
symmetry, the stationary distribution is exponential and does not
depend on the details of a model.  Conversely, when time-reversal
symmetry is broken, the distribution may depend on model details.  The
difference between these two classes of models may be rather subtle.
For example, let us change the multiplicative random exchange from a
fixed fraction of loser's money to a fixed fraction of the total money
of winner and loser.  This modification retains the multiplicative
idea that the amount exchanged is proportional to the amount involved,
but restores time-reversal symmetry and the Boltzmann-Gibbs
distribution.  In the model with $\Delta m=1$ discussed in the next
Section, the difference between time-reversible and time-irreversible
formulations amounts to the difference between impenetrable and
absorbing boundary conditions at $m=0$.  Unlike in physics, in economy
there is no fundamental requirement that interactions have
time-reversal symmetry.  However, in the absence of detailed knowledge
of real microscopic dynamics of economic exchange, the semiuniversal
Boltzmann-Gibbs distribution appears to be a natural starting point.

Moreover, deviations from the Boltzmann-Gibbs law may occur only if
the transition rates $w$ in Eq.\ (\ref{Boltzmann}) explicitly depend
on the agents' money $m$ or $m'$ in an asymmetric manner.  In another
simulation, we randomly preselected winners and losers for every pair
of agents $(i,j)$.  In this case, money flows along directed links
between the agents: $i\!\to\!j\!\to\!k$, and time-reversal symmetry is
strongly broken.  This model is closer to the real economy, in which,
for example, one typically receives money from an employer and pays it
to a grocer, but rarely the reverse.  Nevertheless, we still found the
Boltzmann-Gibbs distribution of money in this model, because the
transition rates $w$ do not explicitly depend on $m$ and $m'$.

\subsection{Nonlinear Boltzmann equation vs. linear master equation}

For the model where agents randomly exchange the constant amount
$\Delta m=1$, the Boltzmann equation is: 
\begin{eqnarray} 
  &&\!\!\!\!\!\!\!\!\!\!\!\!\!\!\!\!
     \frac{dP_m}{dt}=P_{m+1}\sum\limits_{n=0}^\infty P_n
     +P_{m-1}\sum\limits_{n=1}^\infty P_n 
\nonumber \\
  &&{}-P_m\sum\limits_{n=0}^\infty P_n 
 -P_m\sum\limits_{n=1}^\infty P_n 
\label{D0} \\
  &&\!\!\!\!\!\!=(P_{m+1}+P_{m-1}-2P_m)+P_0(P_m-P_{m-1}),
\label{D1}
\end{eqnarray} 
where $P_m\equiv P(m)$ and we have used $\sum_{m=0}^\infty P_m=1$.
The first, diffusion term in Eq.\ (\ref{D1}) is responsible for
broadening of the initial delta-function distribution.  The second
term, proportional to $P_0$, is essential for the Boltzmann-Gibbs
distribution $P_m=e^{-m/T}(1-e^{-1/T})$ to be a stationary solution of
Eq.\ (\ref{D1}).  In a similar model studied in Ref.\ \cite{Redner},
the second term was omitted on the assumption that agents who lost all
money are eliminated: $P_0=0$.  In that case, the total number of
agents is not conserved, and the system never reaches any stationary
distribution.  Time-reversal symmetry is violated, since transitions
into the state $m=0$ are permitted, but not out of this state.

If we treat $P_0$ as a constant, Eq.\ (\ref{D1}) looks like a linear
Fokker-Planck equation \cite{Kinetics} for $P_m$, with the first term
describing diffusion and the second term an external force
proportional to $P_0$.  Similar equations were studied in Ref.\
\cite{Montroll1}.  Eq.\ (\ref{D1}) can be also rewritten as
\begin{equation}
\frac{dP_m}{dt}=P_{m+1}-(2-P_0)P_m+(1-P_0)P_{m-1}.
\label{Fokker-Planck}
\end{equation}
The coefficient $(1-P_0)$ in front of $P_{m-1}$ represents the rate of
increasing money by $\Delta m=1$, and the coefficient 1 in front of
$P_{m+1}$ represents the rate of decreasing money by $\Delta m=-1$.
Since $P_0>0$, the former is smaller than the latter, which results in
the stationary Boltzmann-Gibbs distributions $P_m=(1-P_0)^m$.  An
equation similar to Eq.\ (\ref{Fokker-Planck}) describes a Markov
chain studied for strategic market games in Ref.\ \cite{Shubik-2}.
Naturally, the stationary probability distribution of wealth in that
model was found to be exponential \cite{Shubik-2}.

Even though Eqs.\ (\ref{D1}) and (\ref{Fokker-Planck}) look like
linear equations, nevertheless the Boltzmann equation
(\ref{Boltzmann}) and (\ref{D0}) is a profoundly nonlinear equation.
It contains the product of two probability distribution functions $P$
in the right-hand side, because two agents are involved in money
exchange.  Most studies of wealth distribution \cite{Montroll1} have
the fundamental flaw that they use a single-particle approach.  They
assume that the wealth of an agent may change just by itself and write
a linear master equation for the probability distribution.  Because
only one particle is considered, this approach cannot adequately
incorporate conservation of money.  In reality, an agent can change
money only by interacting with another agent, thus the problem
requires a two-particle probability distribution function.  Using
Boltzmann's molecular chaos hypothesis, the two-particle function is
factorized into a product of two single-particle distributions
functions, which results in the nonlinear Boltzmann equation.
Conservation of money is adequately incorporated in this two-particle
approach, and the universality of the exponential Boltzmann-Gibbs
distribution is transparent.

\subsection{Conclusions}

Everywhere in this chapter we assumed some randomness in the exchange
of money.  Our results would apply the best to the probability
distribution of money in a closed community of gamblers.  In more
traditional economic studies, the agents exchange money not randomly,
but following deterministic strategies, such as maximization of
utility functions \cite{Shubik,Bak}.  The concept of equilibrium in
these studies is similar to mechanical equilibrium in physics, which
is achieved by minimizing energy or maximizing utility.  However, for
big ensembles, statistical equilibrium is a more relevant concept.
When many heterogeneous agents deterministically interact and spend
various amounts of money from very little to very big, the money
exchange is effectively random.  In the future, we would like to
uncover the Boltzmann-Gibbs distribution of money in a simulation of a
big ensemble of economic agents following realistic deterministic
strategies with money conservation taken into account.  That would be
the economics analog of molecular dynamics simulations in physics.
While atoms collide following fully deterministic equations of motion,
their energy exchange is effectively random due to the complexity of
the system and results in the Boltzmann-Gibbs law.

We do not claim that the real economy is in equilibrium.  (Most of the
physical world around us is not in true equilibrium either.)
Nevertheless, the concept of statistical equilibrium is a very useful
reference point for studying nonequilibrium phenomena.

\section{Distribution of income and wealth} 
\label{ch:2}
\subsection{Introduction}
\label{sec:introduction}

In Ch.\ \ref{ch:1} we predicted that the distribution of money should
follow an exponential Boltzmann-Gibbs law.  Unfortunately, we were not
able to find data on the distribution of money.  On the other hand, we
found many sources with data on income distribution for the United
States (USA) and United Kingdom (UK), as well as data on the wealth
distribution in the UK, which are presented in this chapter.  In all
of these data, we find that the great majority of the population is
described by an exponential distribution, whereas the high-end tail
follows a power law.

The study of income distribution has a long history.  Pareto
\cite{Pareto} proposed in 1897 that income distribution obeys a
universal power law valid for all times and countries.  Subsequent
studies have often disputed this conjecture.  In 1935, Shirras
\cite{Shirras} concluded: ``There is indeed no Pareto Law.  It is time
it should be entirely discarded in studies on distribution''.
Mandelbrot \cite{Mandelbrot} proposed a ``weak Pareto law'' applicable
only asymptotically to the high incomes.  In such a form, Pareto's
proposal is useless for describing the great majority of the
population.

Many other distributions of income were proposed: Levy, log-normal,
Champernowne, Gamma, and two other forms by Pareto himself (see a
systematic survey in the World Bank research publication
\cite{Kakwani}).  Theoretical justifications for these proposals form
two schools: socio-economic and statistical.  The former appeals to
economic, political, and demographic factors to explain the
distribution of income (e.\ g.\ \cite{Levy}), whereas the latter
invokes stochastic processes.  Gibrat \cite{Gibrat} proposed in 1931
that income is governed by a multiplicative random process, which
results in a log-normal distribution (see also \cite{Montroll}).
However, Kalecki \cite{Kalecki} pointed out that the width of this
distribution is not stationary, but increases in time.  Levy and
Solomon \cite{Solomon} proposed a cut-off at lower incomes, which
stabilizes the distribution to a power law.

Many researchers tried to deduce the Pareto law from a theory of stochastic
processes.  Gibrat \cite{Gibrat} proposed in 1931 that income and
wealth are governed by multiplicative random processes, which result
in a log-normal distribution.  These ideas were later followed, among
many others, by Montroll and Shlesinger \cite{Montroll}.  However,
already in 1945 Kalecki \cite{Kalecki} pointed out that the log-normal
distribution is not stationary, because its width increases in time.
Modern econophysicists \cite{Solomon,Bouchaud,Sornette} also use
various versions of multiplicative random processes in theoretical
modeling of wealth and income distributions.

Unfortunately, numerous recent papers on this subject do very little
or no comparison at all with real statistical data, much of which is
widely available these days on the Internet.  In order to fill this
gap, we analyzed the data on income distribution in the United States
(US) from the Bureau of Census and the Internal Revenue Service (IRS)
in Ref.\ \cite{DY-income}.  We found that the individual income of
about 95\% of population is described by the exponential law.  The
exponential law, also known in physics as the Boltzmann-Gibbs
distribution, is characteristic for a conserved variable, such as
energy.  In Ref.\ \cite{DY-money}, we argued that, because money
(cash) is conserved, the probability distribution of money should be
exponential.  Wealth can increase or decrease by itself, but money can
only be transferred from one agent to another.  So, wealth is not
conserved, whereas money is.  The difference is the same as the
difference between unrealized and realized capital gains in stock
market.

In Sec.\ \ref{sec:individual}, we propose that the distribution of
individual income is given by an exponential function.  This
conjecture is inspired by the results of Ch.\ \ref{ch:1}, \cite{DY-income},
where we argued that the probability distribution of money in a closed
system of agents is given by the exponential Boltzmann-Gibbs function.
We compare our proposal with the census and tax data for individual
income in USA.  In Sec.\ \ref{sec:pltbc} we show that the exponential
distribution has to be amended for the top 5\% of incomes by a
power-law function.  In Sec.\ \ref{sec:family}, we derive the
distribution function of income for families with two earners and
compare it with census data.  In Sec.\ \ref{sec:wealth}, we discuss
the distribution of wealth.  In Sec.\ \ref{tsallisWealth} we
critically examine several other alternatives proposed in the
literature for the distribution of income and wealth.  Speculations on
the possible origins of the exponential and power-law distribution of
income and conclusions for this chapter are given in Sec.\
\ref{sec:conc2}.

\subsection{Distribution of income for individuals }
\label{sec:individual}

\subsection{Exponential distribution of income}
\label{sec:expInc}

We denote income by the letter $r$ (for ``revenue'').  The probability
distribution function of income, $P(r)$, (called the probability
density in book \cite{Kakwani}) is defined so that the fraction of
individuals with income between $r$ and $r+dr$ is $P(r)\,dr$.  This
function is normalized to unity (100\%): $\int_0^\infty
P(r)\,dr=1$.  We propose that the probability distribution of
individual income is exponential:
\begin{equation}
  P_1(r)=\exp(-r/R)/R,
\label{eq:BG}  
\end{equation}
where the subscript 1 indicates individuals.  Function (\ref{eq:BG})
contains one parameter $R$, equal to the average income: 
$\int_0^\infty r\,P_1(r)\,dr=R$, and analogous to temperature in the
Boltzmann-Gibbs distribution \cite{DY-income}.

\begin{figure}
\centerline{
\epsfig{file=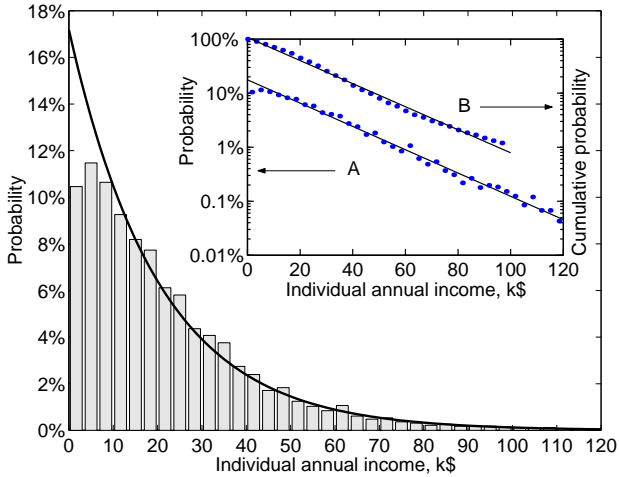, width=0.95\linewidth}}
\caption{ {\small 
  Histogram: Probability distribution of individual income from the
  U.S.\ Census data for 1996 \cite{census}.  Solid line: Fit to the
  exponential law.  Inset plot A: The same with the logarithmic
  vertical scale.  Inset plot B: Cumulative probability distribution
  of individual income from PSID for 1992 \cite{Michigan}.}}
\label{fig:census}
\end{figure}

From the Survey of Income and Program Participation (SIPP)
\cite{census}, we downloaded the variable TPTOINC (total income of a
person for a month) for the first ``wave'' (a four-month period) in
1996.  Then we eliminated the entries with zero income, grouped the
remaining entries into bins of the size 10/3 k\$, counted the numbers
of entries inside each bin, and normalized to the total number of
entries.  The results are shown as the histogram in Fig.\
\ref{fig:census}, where the horizontal scale has been multiplied by 12
to convert monthly income to an annual figure.  The solid line
represents a fit to the exponential function (\ref{eq:BG}).  In the
inset, plot A shows the same data with the logarithmic vertical scale.
The data fall onto a straight line, whose slope gives the parameter
$R$ in Eq.\ (\ref{eq:BG}).  The exponential law is also often written
with the bases 2 and 10:
$P_1(r)\propto2^{-r/R_2}\propto10^{-r/R_{10}}$.  The parameters $R$,
$R_2$ and $R_{10}$ are given in line (c) of Table \ref{tab:R}.

\begin{table}[h]
  \begin{center}
  \tabcolsep 1.1ex
  \begin{tabular}{|c|lccccc|} \hline
& Source & Year & $R$ (\$) & $R_2$ (\$) & $R_{10}$ (\$) & Set size \\  \hline
a & PSID \cite{Michigan} & 1992 & 18,844 & 13,062 & 43,390 & 
  1.39$\times10^3$\\
b & IRS \cite{SailerWeber} & 1993 & 19,686 & 13,645 & 45,329 & 
  1.15$\times10^8$\\
c & SIPP$\rm_p$ \cite{census} & 1996 & 20,286 & 14,061 & 46,710 & 
  2.57$\times10^5$\\
d & SIPP$\rm_f$ \cite{census} & 1996 & 23,242 & 16,110 & 53,517 &
  1.64$\times10^5$\\
e & IRS \cite{Pub1304} & 1997 & 35,200 & 24,399 & 81,051 & 
  1.22$\times10^8$ \\  \hline
  \end{tabular}
  \end{center}
\caption{{\small Parameters $R$, $R_2$, and $R_{10}$ obtained by fitting data
  from different sources to the exponential law (\protect\ref{eq:BG})
  with the bases $e$, 2, and 10, and the sizes of the statistical data
  sets.}}
\label{tab:R}
\end{table}

Plot B in the inset of Fig.\ \ref{fig:census} shows the data from the
Panel Study of Income Dynamics (PSID) conducted by the Institute for
Social Research of the University of Michigan \cite{Michigan}.  We
downloaded the variable V30821 ``Total 1992 labor income'' for
individuals from the Final Release 1993 and processed the data in a
similar manner.  Shown is the cumulative probability distribution of
income $N(r)$ (called the probability distribution in book
\cite{Kakwani}).  It is defined as $N(r)=\int_r^\infty P(r')\, dr'$
and gives the fraction of individuals with income greater than $r$.
For the exponential distribution (\ref{eq:BG}), the cumulative
distribution is also exponential: $N_1(r)=\int_r^\infty P_1(r')\,
dr'=\exp(-r/R)$.  Thus, $R_2$ is the median income; 10\% of population
have income greater than $R_{10}$ and only 1\% greater than $2R_{10}$.
The points in the inset fall onto a straight line in the logarithmic
scale.  The slope is given in line (a) of Table \ref{tab:R}.

\begin{figure}
\centerline{
\epsfig{file=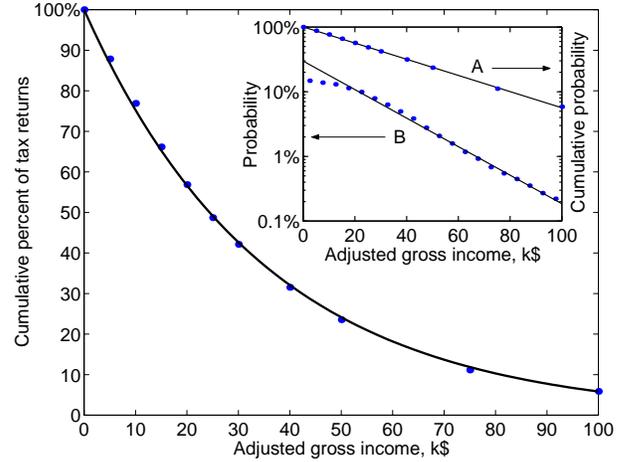,width=0.95\linewidth}}
\caption{{\small 
  Points: Cumulative fraction of tax returns vs income from the IRS
  data for 1997 \cite{Pub1304}.  Solid line: Fit to the exponential
  law.  Inset plot A: The same with the logarithmic vertical scale.
  Inset plot B: Probability distribution of individual income from the
  IRS data for 1993 \cite{SailerWeber}.}}
\label{fig:IRS}
\end{figure}

The points in Fig.\ \ref{fig:IRS} show the cumulative distribution of
tax returns vs income in 1997 from column 1 of Table 1.1 of Ref.\ 
\cite{Pub1304}.  (We merged 1 k\$ bins into 5 k\$ bins in the interval
1--20 k\$.)  The solid line is a fit to the exponential law.  Plot A
in the inset of Fig.\ \ref{fig:IRS} shows the same data with the
logarithmic vertical scale.  The slope is given in line (e) of Table
\ref{tab:R}.  Plot B in the inset of Fig.\ \ref{fig:IRS} shows the
distribution of individual income from tax returns in 1993
\cite{SailerWeber}.  The logarithmic slope is given in line (b) of
Table \ref{tab:R}.

While Figs.\ \ref{fig:census} and \ref{fig:IRS} clearly demostrate the
fit of income distribution to the exponential form, they have the
following drawback.  Their horizontal axes extend to $+\infty$, so the
high-income data are left outside of the plots.  The standard way to
represent the full range of data is the so-called Lorenz curve (for an
introduction to the Lorenz curve and Gini coefficient, see book
\cite{Kakwani}).  The horizontal axis of the Lorenz curve, $x(r)$,
represents the cumulative fraction of population with income below
$r$, and the vertical axis $y(r)$ represents the fraction of income
this population accounts for:
\begin{equation} \label{lorDef}
  x(r)=\int_0^r P(r')\,dr',\quad
  y(r)=\frac{\int_0^r r' P(r')\,dr'}{\int_0^\infty r' P(r')\,dr'}.
\label{eq:xy}
\end{equation}
As $r$ changes from 0 to $\infty$, $x$ and $y$ change from 0 to 1, and
Eq.\ (\ref{eq:xy}) parametrically defines a curve in the
$(x,y)$-space.

Substituting Eq.\ (\ref{eq:BG}) into Eq.\ (\ref{eq:xy}), we find
\begin{equation}
  x(\tilde{r})=1-\exp(-\tilde{r}), \quad
  y(\tilde{r})=x(\tilde{r})-\tilde{r}\exp(-\tilde{r}),
\label{eq:xy1}
\end{equation}
where $\tilde{r}=r/R$.  Excluding $\tilde{r}$, we find the explicit
form of the Lorenz curve for the exponential distribution:
\begin{equation}  \label{iLorenz}
  y=x+(1-x)\ln(1-x).
\label{eq:Lorenz}
\end{equation}
$R$ drops out, so Eq.\ (\ref{eq:Lorenz}) has no fitting parameters.

\begin{figure}
\centerline{
\epsfig{file=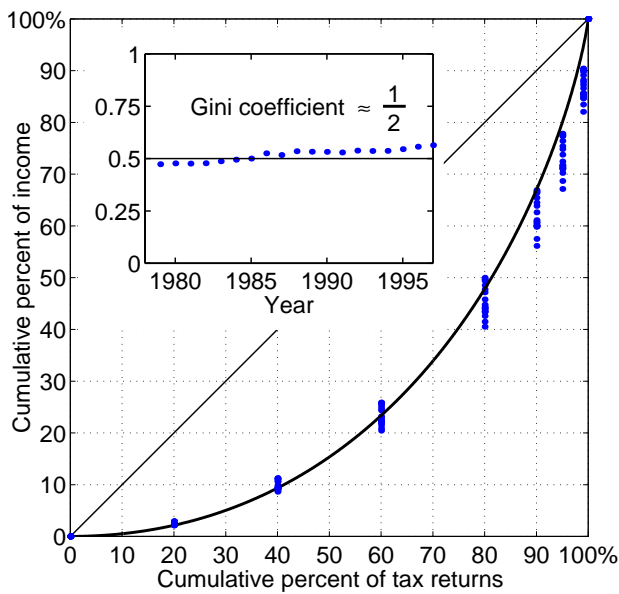,width=0.95\linewidth}}
\caption{{\small
  Solid curve: Lorenz plot for the exponential distribution.  Points:
  IRS data for 1979--1997 \cite{Petska}.  Inset points: Gini
  coefficient data from IRS \cite{Petska}.  Inset line: The calculated
  value 1/2 of the Gini coefficient for the exponential distribution.}}
\label{fig:Lorenz_1}
\end{figure}

The function (\ref{eq:Lorenz}) is shown as the solid curve in Fig.\ 
\ref{fig:Lorenz_1}.  The straight diagonal line represents the Lorenz
curve in the case where all population has equal income.  Inequality
of income distribution is measured by the Gini coefficient $G$, the
ratio of the area between the diagonal and the Lorenz curve to the
area of the triangle beneath the diagonal
\begin{equation}  \label{giniE}
  G=2\int_0^1(x-y)\,dx
\end{equation}
The Gini coefficient is confined between 0 (no inequality) and 1
(extreme inequality).  By substituting Eq.\ (\ref{eq:Lorenz}) into the
integral, we find the Gini coefficient for the exponential
distribution: $G_1=1/2$.

The points in Fig.\ \ref{fig:Lorenz_1} represent the tax data during
1979--1997 from Ref.\ \cite{Petska}.  With the progress of time, the
Lorenz points shifted downward and the Gini coefficient increased from
0.47 to 0.56, which indicates increasing inequality during this
period.  However, overall the Gini coefficient is close to the value
0.5 calculated for the exponential distribution, as shown in the inset
of Fig.\ \ref{fig:Lorenz_1}.

\subsection{Power-law tail and ``Bose'' condensation}
\label{sec:pltbc}

As Fig.\ \ref{fig:Lorenz_1} shows, the Lorenz curve deviates from the
theoretical Lorenz curve implied by the exponential distribution,
mostly for the top 20\% of tax returns.  Moreover, as explained in
the previous section, the Lorenz curve for a pure exponential
distribution is independent of temperature (the scale of the
distribution).  Therefore the variations in the Lorenz curves over the
period 1979-1997 suggest that the shape of the distribution changes.

In Sec.\ \ref{sec:introduction} we mentioned the early proposal of
Pareto, who claimed that the income distribution obeys a universal
power law valid for all times and countries \cite{Pareto}.  The Pareto
distribution $P(r)=A/r^\alpha$ has several undesirable properties.
It diverges for low incomes, and if $\alpha<2$ the distribution
has a diverging first moment.  These two properties are never found in
real data.  Moreover, our study has shown that at least for the
overwhelming majority (95\%) of the population the distribution of
income is exponential. 

The data for the remaining top 5\% percent of the population is hard
to get and often it is unreliable.  Because the number of people with
income in the top few percents is small, census data is poor.  The
Internal Revenue Service (IRS) has reliable data but it seldom reports
it.  We managed to find a data set from IRS which contains high income
data
\cite{Pub1304}.

\begin{figure}
\centerline{
\epsfig{file=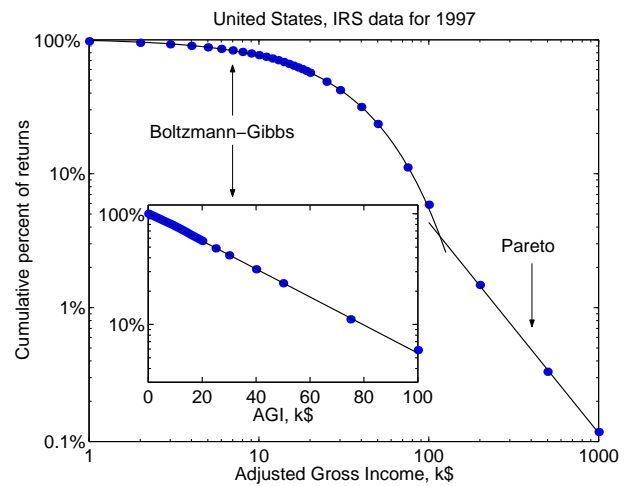,width=0.95\linewidth}}
\caption{{\small
  Cumulative probability distribution of US individual
  income for 1997 in log-log scale, with points (raw data) and solid
  lines (exponential and power-law fit).  The inset shows the
  exponential regime and the fit with a Boltzmann-Gibbs distribution
  in the log-linear scale. }}
\label{fig:irs97G+P_left}
\end{figure}

Fig.\ \ref{fig:irs97G+P_left} shows United States IRS income data for 1997
\cite{Pub1304}.  The data points represent the cumulative probability 
distribution in log-log scale.  Although for large income values
$r>120$k\$ we have only three data points, the points align on a
straight line in accordance with a Pareto power-law distribution.

From the plot it is evident that there is a discontinuity around
120k\$ where the exponential and power-law regimes intersect.  We
conclude that it is not possible to describe the entire income
distribution with one single differentiable function.  The two regimes
of the probability distribution are clearly separated and this may be
due to different income dynamics in the two regimes.  Among others,
the Adjusted Gross Income contains capital gains that is stock-market
gains/losses.  It is conceivable that for the top 5\% of the
population, capital gains rather than the labor income account for
the majority share of the total income.  The capital gains
contribution to the Adjusted Growth Income may be responsible for the
observed power-law in the tail of the income distribution.

\begin{figure}
\centerline{
  \epsfig{file=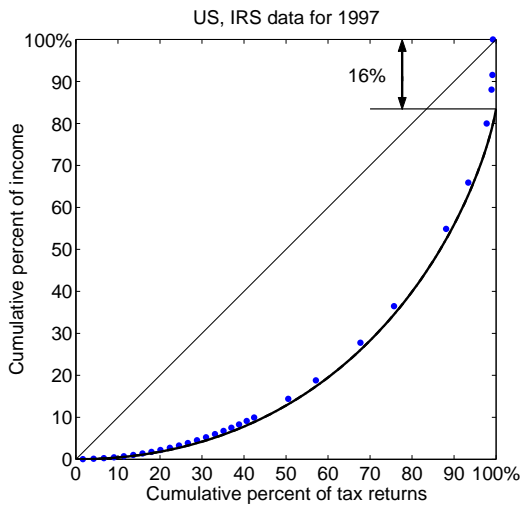,width=0.80\linewidth}}
\caption{ {\small 
  Lorenz plot with points (raw data) and solid line
  (function (\ref{iLorenzBose}) with fraction $b=16\%$).}}
\label{fig:irs97G+P_right}
\end{figure}

We plot the Lorenz curve for income in Fig.\ \ref{fig:irs97G+P_right}.
As in Sec.\ \ref{sec:individual} the horizontal and vertical
coordinates are the cumulative population $x(r)$ and the cumulative
income $y(r)$ from (\ref{lorDef}).  An imaginary line through the data
shows an abrupt change in derivative for the last 2\% of the tax
returns.  This sudden change in derivative around the 98\% mark on the
horizontal axis of Fig.\ \ref{fig:irs97G+P_right} gives an almost infinite
slope for the Lorenz curve.  Physically, an infinite slope in the
Lorenz function for arguments $x$ just slightly less than one, is
equivalent to having a finite amount of total society's income in the
hand of a very few individuals. We have coined for this effect the
name ``Bose condensation of income'' because of similarities with the
celebrated phase-transition from statistical mechanics.

To understand quantitatively the ``Bose condensation'' of income, we
calculate $f$ the ratio of the total income of the society if all the
population is described by the exponential law, to the actual total
income of the population.  For the USA tax income data for 1997 shown
in Fig.\ \ref{fig:irs97G+P_right} this fraction was
$f=0.84$, which means that the ``condensate'' has $b=1-f=16\%$ of the
total income of the population.  The analytical formula for the Lorenz
curve in this case is
\begin{equation} \label{iLorenzBose}
   y=(1-b)[x + (1-x)\ln(1-x)] + b\delta(1-x), 
\end{equation}
where the first term represents the contribution of the
Boltzmann-Gibbs exponential regime and the second term represents the
contribution of the Pareto power-law tail.  It is remarkable that as
in the case of (\ref{iLorenz}), Eq.\ \ref{iLorenzBose} does not have
any fitting parameters.  The condensate fraction $b$ is completely
determined by temperature, which in turn is determined from the
probability distribution.

The function (\ref{iLorenzBose}) is plotted as a solid line in 
Fig.\ \ref{fig:irs97G+P_left}.  One can see that the data
systematically deviate from the exponential law because of the income
concentrated in the power-law tail; however, the deviation is not very
big.  The inequality of the US income distribution was also increasing
during that time period \cite{DY-income}, which implies that the 
size of the ``Bose condensate'' has increased in time, too.

The Gini coefficient defined in (\ref{giniE}) can be calculated for
the Lorenz curve given by (\ref{iLorenzBose}).  The effect of the
power-law tails changes the value of the Gini coefficient from
$G_1=1/2$ in the case of a pure exponential distribution to
$G_b=(1+b)/2$.

\subsection{Geographical variations in income distribution}
\label{geo}

In the previous sections we found that the distribution of income is
an exponential for the large majority of the population followed by a
power-law tail for the top few percent of the population.  It would be
interesting to establish the universality of this shape for various
other countries.

\begin{figure}
\centerline{
  \epsfig{file=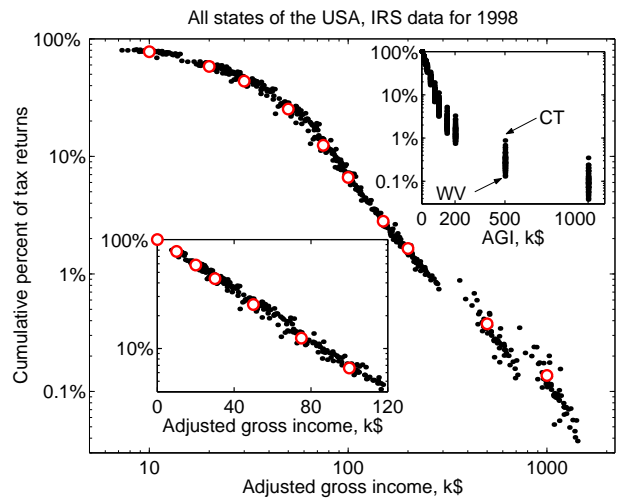,width=0.95\linewidth}}
\caption{{\small Cumulative probability distributions of
 yearly individual income for different states of the USA shown as raw
 data (top inset) and scaled data in log-log, log-linear
 (lower inset).}}
\label{fig:USincome_left}
\end{figure}

We obtained the data on distribution of the yearly individual income
in 1998 for each of the 50 states and the District of Columbia that
constitute the USA from the Web site of the IRS \cite{IRSsite}.  We
plot the original raw data for the cumulative distribution of income
in log-linear scale in the upper inset of Fig.\
\ref{fig:USincome_left}.  The points spread significantly,
particularly at higher incomes.  For example, the fraction of
individuals with income greater than 1~M\$ varies by an order of
magnitude between different states.  However, after we rescale the
data in the manner described in the preceding section, the points
collapse on a single curve shown in log-log scale in the main panel
and log-linear scale in the lower inset.  The open circles represent
the US average, obtained by treating the combined data for all states
as a single set.  We observe that the distribution of higher incomes
approximately follows a power law with the exponent
$\alpha=1.7\pm0.1$, where the $\pm0.1$ variation includes 70\% of all
states.  On the other hand, for about 95\% of individuals with lower
incomes, the distribution follows an exponential law with the average
US temperature $R_{US}=36.4$~k\$ ($R_{US}^{(2)}=25.3$~k\$ and
$R_{US}^{(10)}=83.9$~k\$).  The temperatures of the individual states
differ from $R_{US}$ by the amounts shown in Table \ref{table}.  For
example, the temperature of Connecticut (CT) is 1.25 times higher and
the temperature of West Virginia (WV) is 0.78 times lower than the
average US temperature.

\begin{figure}
\centerline{
  \epsfig{file=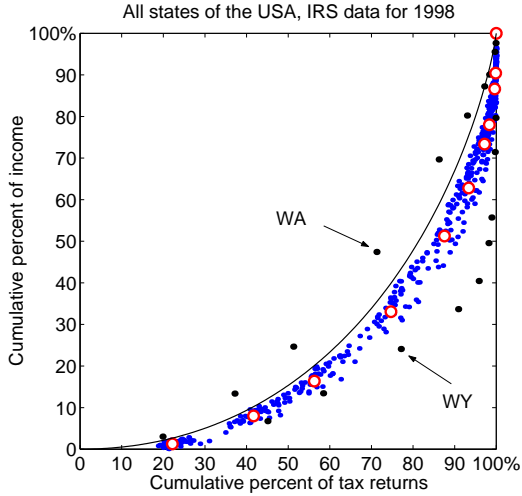,width=0.80\linewidth}}
\caption{{\small  Lorenz plot with points (raw data) and solid line 
 (function (\ref{iLorenz}) calculated for the exponential law).}}
\label{fig:USincome_right}
\end{figure}

\begin{table}
\caption{Deviations of the state temperatures from the average US
temperature.}
\scriptsize
\begin{tabular}{@{\extracolsep{-0.0em}}*{11}{c}@{\extracolsep{-0.0em}}}
\hline
 CT  &  NJ  &  MA  &  MD  &  VA  &  CA  &  NY  &  IL  &  CO  &  NH  &  

 AK  \\ [-0.5ex]

25\% & 24\% & 14\% & 14\% &  9\% &  9\% &  7\% & 6\%  &  6\% &  5\% &   

 5\%  \\ \hline
 DC  &  DE  &  MI  &  WA  &  MN  &  GA  &  TX  &  RI  &  AZ  &  PA  & 

 FL  \\ [-0.5ex]

 5\% &  4\% &  4\% &  2\% &  1\% &  0\% & -1\% & -3\% & -3\% & -3\% & 

-4\%  \\ \hline
 KS  &  OR  &  HI  &  NV  &  NC  &  WI  &  IN  &  UT  &  MO  &  VT  &  

 TN    \\[-0.5ex]

-5\% & -6\% & -7\% & -7\% & -7\% & -8\% & -8\% & -9\% & -9\% & -9\% & 

-11\%   \\ \hline
 NE  &  OH  &  LA  &  AL  &  SC  &  IA  &  WY  &  NM  &  KY  & ID  & 

 OK    \\[-0.5ex]

-12\% & -12\% & -13\% & -13\% & -13\% & -14\% & -14\% & -14\% & -14\% & 

-15\% &  -16\%  \\ \hline
  ME   &  MT   &  AR   &  SD   &  ND   &  MS   &  WV  &  &  &  &\\[-0.5ex] 

 -16\% & -19\% & -19\% & -20\% & -20\% & -21\% & -22\% &  &  &  &\\ \hline
\end{tabular}
\label{table}
\end{table}

The Lorenz plot for all states is shown in Fig.\
\ref{fig:USincome_right} together with the solid curve representing
Eq.\ (\ref{iLorenz}).  The majority of points are well clustered and
are not too far from the solid curve.  The exceptions are Wyoming (WY)
with much higher inequality and the Washington state (WA) with
noticeably lower inequality of income distribution.  The average US
data, shown by open circles, is consistent with our previous results
\cite{DY-income}.  Unlike in the UK case, we did not make any
adjustment in the US case for individuals with income below the
threshold, which appears to be sufficiently low.

\begin{figure}
\centerline{
  \epsfig{file=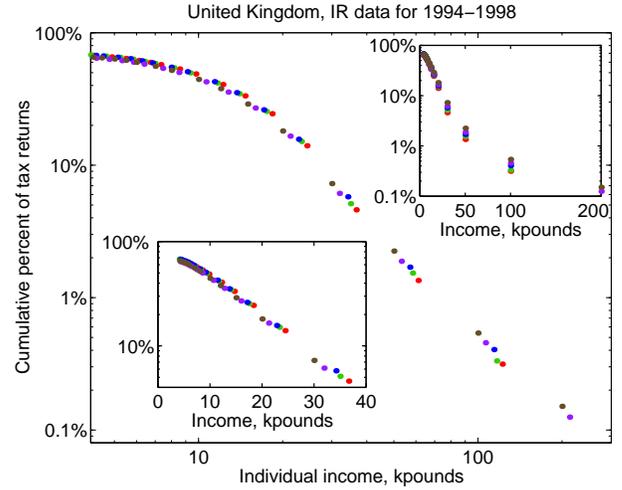,width=0.95\linewidth}}
\caption{{\small Cumulative probability distributions of
 yearly individual income in the UK shown as raw data (top inset) and
 scaled data in log-log (left panel), log-linear (lower inset), and
 Lorenz (right panel) coordinates.  Solid curve: fit to function
 (\ref{iLorenz}) calculated for the exponential law.}}
\label{fig:UKincome_left}
\end{figure}

We obtained the data on the yearly income distribution in the UK for
1997/8 and 1998/9 from the Web site of the IR \cite{UKincome}.  The
data for 1994/5, 1995/6, and 1996/7 were taken from the Annual
Abstract of Statistics derived from the IR \cite{ONS}.  The data for
these 5 years are presented graphically in Fig.\ \ref{fig:UKincome_left}.
In the upper inset, the original raw data for the
cumulative distribution are plotted in log-linear scale.  For not too
high incomes, the points form straight lines, which implies the
exponential distribution $N(r)\propto\exp(-r/R)$, where $r$ stands for
income (revenue), and $R$ is the income ``temperature''.  However, the
slopes of these lines are different for different years.  The
temperatures for the years 1994/5, 1995/6, 1996/7, and 1997/8 differ
from the temperature for 1998/9, $R_{UK}^{(98/9)}=11.7$~k$\pounds$
($R_{UK}^{(2)}=8.1$~k$\pounds$ and $R_{UK}^{(10)}=26.9$~k$\pounds$),
by the factors 0.903, 0.935, 0.954, and 0.943.  To compensate for this
effect, we rescale the data.  We divide the horizontal coordinates
(incomes) of the data sets for different years by the quoted above
factors and plot the results in log-log scale in the main panel and
log-linear scale in the lower inset.  We observe scaling: the collapse
of points onto a single curve.  Thus, the distributions $N_i(r)$ for
different years $i$ are described by a single function $f(r/R_i)$.
The main panel shows that this scaling function $f$ follows a power
law with the exponent $\alpha=2.0$-2.3 at high incomes.  The lower
inset shows that $f$ has an exponential form for about 95\% of
individuals with lower incomes.  These results qualitatively agree
with a similar study by Cranshaw \cite{Cranshaw}.  He proposed that
the $P(r)$ data for lower incomes are better fitted by the Gamma
distribution $\Gamma(r)\propto r^\beta\exp(-r/R)$.  For simplicity, we
chose not to introduce the additional fitting parameter $\beta$.

We must mention that the individuals with income below a certain
threshold are not required to report to the IR.  That is why the data
in the lower inset do not extend to zero income.  We extrapolate the
straight line to zero income and take the intercept with the vertical
axis as 100\% of individuals.  Thus, we imply that the IR data does
not account for about 25-27\% of individuals with income below the
threshold.

\begin{figure}
\centerline{
  \epsfig{file=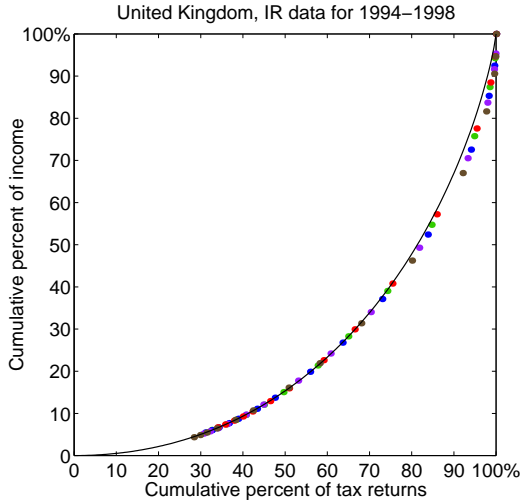,width=0.80\linewidth}}
\caption{{\small Cumulative probability distributions of
 yearly individual income in the UK shown as raw data (top inset) and
 scaled data in log-log (left panel), log-linear (lower inset), and
 Lorenz (right panel) coordinates.  Solid curve: fit to function
 (\ref{iLorenz}) calculated for the exponential law.}}
\label{fig:UKincome_right}
\end{figure}

The Lorenz curve for the distribution of the UK income is shown in
Fig.\ \ref{fig:UKincome_right}.  We treat the number of individuals
below the income threshold and their total income as adjustable
parameters, which are the horizontal and vertical offsets of the
coordinates origin relative to the lowest known data point.  These
parameters are chosen to fit the Lorenz curve for the exponential law
(\ref{iLorenz}) shown as the solid line.  The fit is very good, and is
illustrated in Fig.\ \ref{fig:UKincome_left}.  The horizontal offsets
are 28-34\%, which is roughly consistent with the numbers quoted for
the lower inset of the left panel.

The income temperature for the UK in 1998/9 was
$R_{UK}=11.7$~k$\pounds$ and for the US in 1998 was $R_{US}=36.4$~k\$.
Using the exchange rate as of December 31, 1998 to convert pounds into
dollars \cite{currency}, we find that the UK temperature was
$R_{UK}=19.5$~k\$, which is 1.87 times lower than the US temperature.
The difference in temperatures indicates nonequilibrium, which can be
exploited to create a thermal machine \cite{DY-money}.  The gain
(profit) produced by such a thermal machine is proportional to the
difference in temperatures.  In agreement with the second law of
thermodynamics, money would flow from a high-temperature system to a
low-temperature one.  This may explain the huge trade deficit of the
USA in global international trade with other, lower-temperature
countries.  The variation of temperatures between different states of
the USA is shown in Table \ref{table}.

\subsection[Distribution of income for two-earner families]
{Distribution of income for \hfill \\ families}
\label{sec:family}

\begin{figure}
\centerline{
\epsfig{file=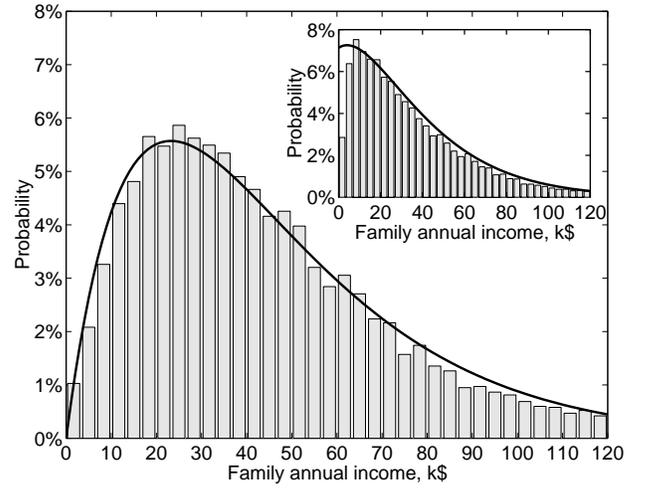,width=0.95\linewidth}}
\caption{{\small 
  Histogram: Probability distribution of income for families with two
  adults in 1996 \cite{census}.  Solid line: Fit to Eq.\ 
  (\ref{eq:family}).  Inset histogram: Probability distribution of
  income for all families in 1996 \cite{census}. Inset solid line:
  $0.45P_1(r)+0.55P_2(r)$.}}
\label{fig:census_2}
\end{figure}

Now let us discuss the distribution of income for families with two
earners.  The family income $r$ is the sum of two individual incomes:
$r=r_1+r_2$.  Thus, the probability distribution of the family income
is given by the convolution of the individual probability
distributions \cite{Feller}.  If the latter are given by the
exponential function (\ref{eq:BG}), the two-earners probability
distribution function $P_2(r)$ is
\begin{equation}
  P_2(r)=\int_0^{r}P_1(r')P_1(r-r')\,dr'= \frac{r}{R^2}e^{-r/R}.
\label{eq:family}
\end{equation}
The function $P_2(r)$ (\ref{eq:family}) differs from the function
$P_1(r)$ (\ref{eq:BG}) by the prefactor $r/R$, which reflects the
phase space available to compose a given total income out of two
individual ones.  It is shown as the solid curve in Fig.\ 
\ref{fig:census_2}.  Unlike $P_1(r)$, which has a maximum at zero
income, $P_2(r)$ has a maximum at $r=R$ and looks qualitatively
similar to the family income distribution curves in literature
\cite{Levy}.

From the same 1996 SIPP that we used in Sec.\ \ref{sec:individual}
\cite{census}, we downloaded the variable TFTOTINC (the total family
income for a month), which we then multiplied by 12 to get annual
income.  Using the number of family members (the variable EFNP) and
the number of children under 18 (the variable RFNKIDS), we selected
the families with two adults.  Their distribution of family income is
shown by the histogram in Fig.\ \ref{fig:census_2}.  The fit to the
function (\ref{eq:family}), shown by the solid line, gives the
parameter $R$ listed in line (d) of Table \ref{tab:R}.  The families
with two adults and more than two adults constitute 44\% and 11\% of
all families in the studied set of data.  The remaining 45\% are the
families with one adult.  Assuming that these two classes of families
have two and one earners, we expect the income distribution for all
families to be given by the superposition of Eqs.\ (\ref{eq:BG}) and
(\ref{eq:family}): $0.45P_1(r)+0.55P_2(r)$.  It is shown by the solid
line in the inset of Fig.\ \ref{fig:census_2} (with $R$ from line (d)
of Table \ref{tab:R}) with the all families data histogram.

\begin{figure}
\centerline{
\epsfig{file=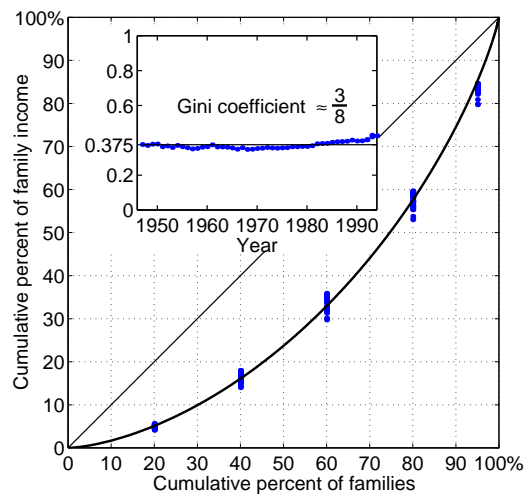,width=0.80\linewidth}}
\caption{{\small 
  Solid curve: Lorenz plot (\protect\ref{eq:xy2}) for distribution
  (\protect\ref{eq:family}).  Points: Census data for families,
  1947--1994 \cite{history}.  Inset points: Gini coefficient data for
  families from Census \cite{history}.  Inset line: The calculated
  value 3/8 of the Gini coefficient for distribution
  (\protect\ref{eq:family}).}}
\label{fig:Lorenz_2}
\end{figure}

By substituting Eq.\ (\ref{eq:family}) into Eq.\ (\ref{eq:xy}), we
calculate the Lorenz curve for two-earners families:
\begin{eqnarray}
  x(\tilde{r}) &=& 1 - (1+\tilde{r}) e^{-\tilde{r}}, \\
  y(\tilde{r}) &=& x(\tilde{r}) -\tilde{r}^2 e^{-\tilde{r}}/2.
\label{eq:xy2}
\end{eqnarray}
It is shown by the solid curve in Fig.\ \ref{fig:Lorenz_2}.  Given
that $x-y=\tilde{r}^2\exp(-\tilde{r})/2$ and
$dx=\tilde{r}\exp(-\tilde{r})\,d\tilde{r}$, the Gini coefficient for
two-earners families is: $G_2=2\int_0^1(x-y)\,dx=
\int_0^\infty\tilde{r}^3\exp(-2\tilde{r})\,d\tilde{r}=3/8=0.375$.  The
points in Fig.\ \ref{fig:Lorenz_2} show the Lorenz data and Gini
coefficient for family income during 1947--1994 from Table 1 of Ref.\
\cite{history}.  The Gini coefficient is very close to the calculated
value 0.375.

\begin{figure}
\centerline{
\epsfig{file=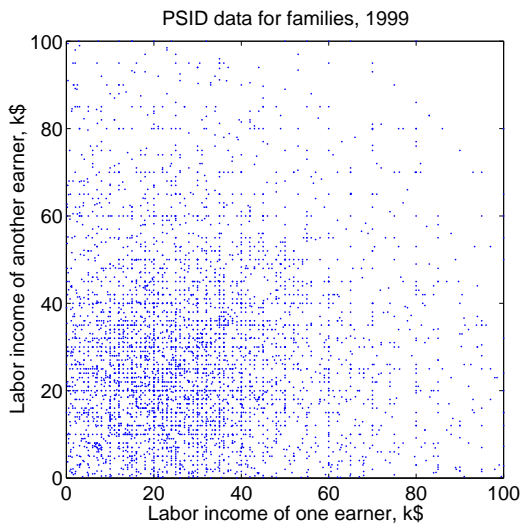,width=0.80\linewidth}}
\caption{{\small 
  Points: The income for families with two earners in 1999 
  \cite{Michigan}.  One family is represented by two points 
  $(r_1,r_2)$ and $(r_2,r_1)$.  The existence of correlations 
  between $r_1$ and $r_2$ is equivalent to a clustering of points 
  along the diagonal.  The data shows little evidence for such 
  correlation.}}
\label{fig:twoEarnersPSID}
\end{figure}

The fundamental assumption which underlies the derivation of
(\ref{eq:family}) is the independence of the two incomes $r_1$ and
$r_2$ of the two-earner family. While the good fit of the function
(\ref{eq:family}) is an implicit validation of this assumption, it
would be good to show that the assumption holds true directly from
data.  The Census database does not provide the individual values of
$r_1$ and $r_2$, but only their sum $r=r_1+r_2$.  We found that the
PSID survey \cite{Michigan} does have give for a family with two
earners, each earners contribution to the total income of the family.
We downloaded the variable ER16463 (total labor income, head) and
ER16465 (total labor income, wife).  We represent in Fig.\
\ref{fig:twoEarnersPSID} each family by two points $(r_1,r_2)$ and
$(r_2,r_1)$.  The existence of correlations between $r_1$ and $r_2$ is
equivalent to a clustering of points along the diagonal.  Or stated
otherwise, the independence of $r_1$ and $r_2$ should result in a
uniform distribution of points on infinitesimal strips perpendicular
to the diagonal. The plotted data shows little, if any evidence for
such correlation.

The distributions of the individual and family income differ
qualitatively.  The former monotonically increases toward the low end
and has a maximum at zero income (Fig.\ \ref{fig:census}).  The
latter, typically being a sum of two individual incomes, has a maximum
at a finite income and vanishes at zero (Fig.\ \ref{fig:census_2}).
Thus, the inequality of the family income distribution is smaller.
The Lorenz data for families follow the different Eq.\ (\ref{eq:xy2}),
again without fitting parameters, and the Gini coefficient is close to
the smaller calculated value 0.375 (Fig.\ \ref{fig:Lorenz_2}).
Despite different definitions of income by different agencies, the
parameters extracted from the fits (Table \ref{tab:R}) are consistent,
except for line (e).

The qualitative difference between the individual and family income
distributions was emphasized in Ref.\ \cite{SailerWeber}, which split
up joint tax returns of families into individual incomes and combined
separately filed tax returns of married couples into family incomes.
However, Refs.\ \cite{Pub1304} and \cite{Petska} counted only
``individual tax returns'', which also include joint tax returns.
Since only a fraction of families file jointly, we assume that the
latter contribution is small enough not to distort the tax returns
distribution from the individual income distribution significantly.
Similarly, the definition of a family for the data shown in the inset
of Fig.\ \ref{fig:census_2} includes single adults and one-adult
families with children, which constitute 35\% and 10\% of all
families.  The former category is excluded from the definition of a
family for the data \cite{history} shown in Fig.\ \ref{fig:Lorenz_2},
but the latter is included.  Because the latter contribution is
relatively small, we expect the family data in Fig.\
\ref{fig:Lorenz_2} to approximately represent the two-earners
distribution (\ref{eq:family}).  Technically, even for the families
with two (or more) adults shown in Fig.\ \ref{fig:census_2}, we do not
know the exact number of earners.

With all these complications, one should not expect perfect accuracy
for our fits.  There are deviations around zero income in Figs.\
\ref{fig:census}, \ref{fig:IRS}, and \ref{fig:census_2}.  The fits
could be improved there by multiplying the exponential function by a
polynomial.  However, the data may not be accurate at the low end
because of underreporting.  For example, filing a tax return is not
required for incomes below a certain threshold, which ranged in 1999
from \$2,750 to \$14,400 \cite{Pub1040}.  As the Lorenz curves in
Figs.\ \ref{fig:Lorenz_1} and \ref{fig:Lorenz_2} show, there are also
deviations at the high end, possibly where Pareto's power law is
supposed to work.  Nevertheless, the exponential law gives an overall
good description of income distribution for the great majority of the
population.

Figs.\ \ref{fig:census} and \ref{fig:IRS} demonstrate that the
exponential law (\ref{eq:BG}) fits the individual income distribution
very well.  The Lorenz data for the individual income follow Eq.\
(\ref{eq:Lorenz}) without fitting parameters, and the Gini coefficient
is close to the calculated value 0.5 (Fig.\ \ref{fig:Lorenz_1}). 

\begin{figure}
\centerline{
\epsfig{file=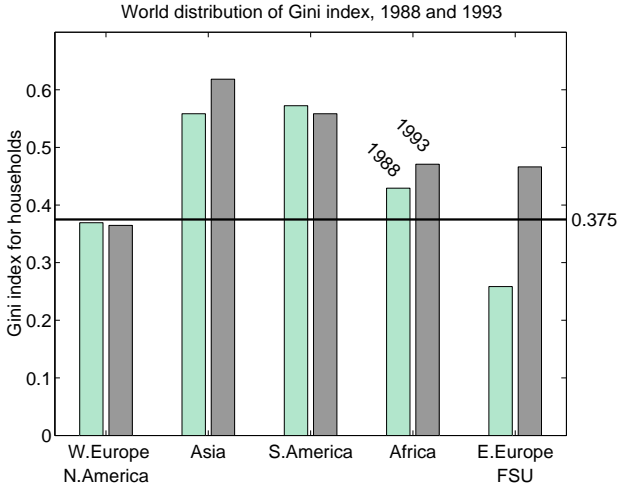,width=0.95\linewidth}}
\caption{{\small 
  Distribution of Gini index for households across the globe for two
  different years, 1988 and 1993 \cite{Milanovic}.  For West Europe
  and North America, the Gini index is close to 0.375, the value
  predicted in Sec.\ \ref{sec:family}.  A sharp increase in inequality
  is observed in the Eastern Europe and the Former Soviet Union (FSU)
  republics after the fall of the ``iron curtain'' when socialist
  economies were replaced by market-like economies. }}
\label{fig:giniWB}
\end{figure}

It is interesting to study the distribution of the Gini index across
the globe.  We present in Fig.\ \ref{fig:giniWB} data from a World
Bank publication \cite{Milanovic}.  This is an question of utmost
importance for the World Bank since the distribution of the Gini index
reflects the degree of inequality (poverty) in various regions of the
globe.  For West Europe and North America, the Gini index is close to
0.375, the value we predicted in Sec.\ \ref{sec:family}.  A sharp
increase in inequality is observed in the Eastern Europe and the
Former Soviet Union (FSU) republics after the fall of the ``iron
curtain'' when socialist economies where replaced by market-like
economies.  We conjecture that the high value of the Gini index in
Asia may be due to the fact that a typical family in Asia has only one
earner, so the Gini index is close to a value of 0.5 as observed for
individuals.

As the evidence from Fig.\ \ref{fig:Lorenz_2} and Fig.\
\ref{fig:giniWB} shows the value of the Gini index varies in time 
and across the globe, but for the Western world it is close to the
value of 0.375, as predicted in Sec.\ \ref{sec:family}.

\subsection{Distribution of wealth}
\label{sec:wealth}

In this section, we discuss the cumulative probability distribution of
wealth $N(w)$=(the number of people whose individual wealth is greater
than $w$)/(the total number of people).  A plot of $N$ vs.\ $w$ is
equivalent to a plot of a person's rank in wealth vs. wealth, which is
often used for top richest people \cite{Solomon97}.  We will use the
power law, $N(w)\propto 1/w^\alpha$, and the exponential law
$N(w)\propto\exp(-w/W)$, to fit the data.  These distributions are
characterized by the exponent $\alpha$ and the ``temperature'' $W$.
The corresponding probability densities, $P(w)=-dN(w)/dw$, also follow
a power law or an exponential law.  For the exponential law, it is
also useful to define the temperatures $W^{(2)}$ (also known as the
median) and $W^{(10)}$ using the bases of 1/2 and 1/10:
$N(w)\propto(1/2)^{w/W^{(2)}}\propto(1/10)^{w/W^{(10)}}$.

\begin{figure}
\centerline{
  \epsfig{file=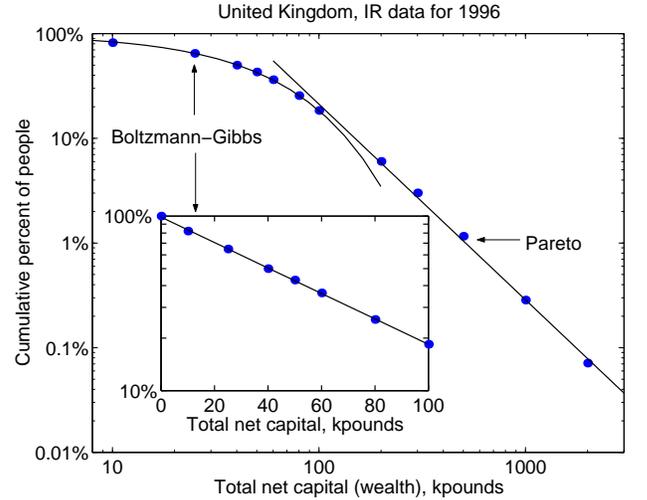,width=0.95\linewidth}}
\caption{{\small Left panel: Cumulative probability distribution of total
  net capital (wealth) shown in log-log, log-linear (inset)
  coordinates.  Points: the actual data.  Solid lines: fits to the
  exponential (Boltzmann-Gibbs) and power (Pareto) laws.}}
\label{fig:UKwealth_left}
\end{figure}

\begin{figure}
\centerline{
   \epsfig{file=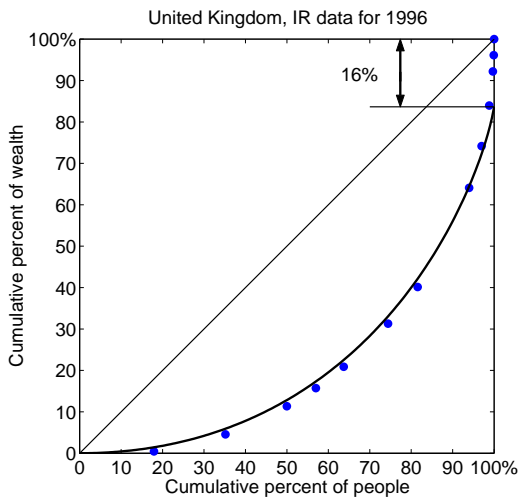,width=0.80\linewidth}}
\caption{{\small 
  Points: Lorenz plot for wealth, United Kingdom 1996 \cite{UKwealth}.
  Solid curve: The Lorenz curve given by (\ref{wealthLorenz}) with 
  a condensate fraction of $b=16\%$.}}
\label{fig:UKwealth_right}
\end{figure}

The distribution of wealth is not easy to measure, because people do
not report their total wealth routinely.  However, when a person dies,
all assets must be reported for the purpose of inheritance tax.  Using
these data and an adjustment procedure, the British tax agency, the
Inland Revenue (IR), reconstructed wealth distribution of the whole UK
population.  In Figs.\ \ref{fig:UKwealth_left} and 
\ref{fig:UKwealth_right}, we present the 1996 data
obtained from their Web site \cite{UKwealth}.  Fig.\ 
\ref{fig:UKwealth_left} shows the cumulative probability as a function 
of the personal total net capital (wealth), which is composed of
assets (cash, stocks, property, house\-hold goods, etc.)  and
liabilities (mortgages and other debts).  The main panel illustrates
in the log-log scale that above 100~k$\pounds$ the data follow a power
law with the exponent $\alpha=1.9$.  The inset shows in the log-linear
scale that below 100~k$\pounds$ the data is very well fitted by an
exponential distribution with the temperature $W_{UK}=59.6$~k$\pounds$
($W_{UK}^{(2)}=41.3$~k$\pounds$ and $W_{UK}^{(10)}=137.2$~k$\pounds$).

Since we have estabilished that the distribution of wealth has an
exponential regime followed by a power-law tail, we plot the Lorenz
curve for wealth in the right panel of Fig.\ \ref{fig:UKwealth_right}.  As
in Sec.\ \ref{sec:individual} the horizontal and vertical coordinates
are the cumulative population $x(w)$ and the cumulative wealth $y(w)$:
\begin{eqnarray}
&& \!\!\!\!\!\!\!\!\!\! x(w)=\int_0^wP(w')\,dw', \\
&& \!\!\!\!\!\!\!\!\!\! 
   y(w)=\int_0^w w'P(w')\,dw'\,/\,\int_0^\infty w'P(w')\,dw'.
\end{eqnarray}

As in the case of US income data Fig.\ \ref{fig:irs97G+P_right}, there
is an abrupt change in the Lorenz curve for the top few percent of
people.  We interpret this fact in the same way as in Sec.\
\ref{sec:pltbc}, as a ``Bose condensation of wealth''.  Again, we
calculate $f$ the fraction of the total wealth of the society if all
people are described by the exponential law, to the actual total
wealth of the society.  For the United Kingdom in 1996, this fraction
was $f=0.84$, which means that the ``condensate'' has $b=1-f=16\%$ of
the total wealth of the system. The functional form for the Lorenz
curve for wealth is
\begin{equation} \label{wealthLorenz}
   y=(1-b)[x + (1-x)\ln(1-x)] + b\delta(1-x).
\end{equation}
This function is plotted as a solid line in Fig.\
\ref{fig:UKwealth_right}.  One can see that the data systematically
deviate from the exponential law because of the wealth concentrated in
the power-law tail; however, the deviation is not very big.  The
so-called Gini coefficient \cite{Kakwani,DY-income}, which measures
the inequality of wealth distribution, has increased from 64\% in 1984
to 68\% in 1996 \cite{UKwealth}.  This value is bigger than the Gini
coefficient 50\% for a purely exponential distribution
\cite{DY-income}. 

While there seems to be no controversy about the fact that a few
people hold a finite fraction of the entire wealth, the specific value
of this fraction varies widely in the literature.  Fractions as high
as 80-90\% were reported in the literature \cite{Anderson} without any
support from data.  We provide a clear procedure of how to calculate
this factor, and we find that $b=16\%$.

\subsection[Other proposed distributions for income and wealth]
{Other distributions for income and wealth}
\label{tsallisWealth}

As we discussed in the introduction, Sec.\ \ref{sec:introduction}
there have been many proposed functional forms for the distribution of
income and wealth.  In this section we will investigate several of
them.  

In Secs.\ \ref{sec:individual}, \ref{sec:pltbc}, \ref{geo} and
\ref{sec:wealth} we have shown ample evidence to support our 
findings that the distribution on income and wealth have a similar
structure with an exponential part followed by a power-law tail.
These results immediately invalidate the proposed Pareto or Gamma
distributions as the correct distribution of income.

\begin{figure}
\centerline{
\epsfig{file=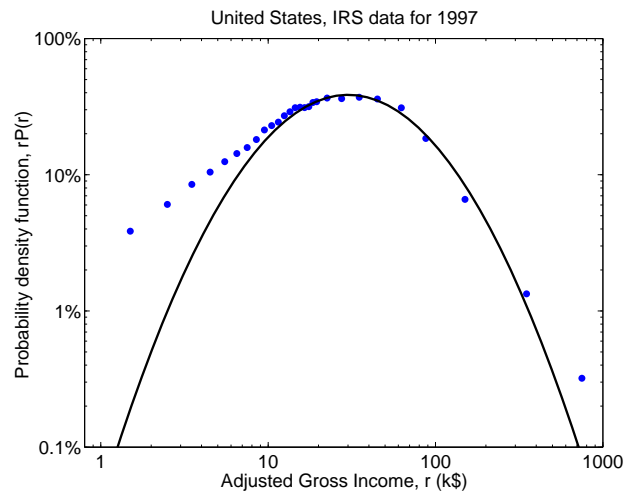,width=0.95\linewidth}}
\caption{ {\small 
  Points: Calculated probability density $P(r)$ for United States IRS
  income data for 1997 multiplied by income $r$, i.e. $rP(r)$.  Curve:
  Fit of data with function $\tilde{P}(r)=rP_{LN}(r)$
  (\ref{lognDistr}) with a quadratic function.  Clearly, the lognormal
  distribution fails to describe the data points accurately.}}
\label{fig:ln97}
\end{figure}

Another popular distribution which has been proposed for many years as
the distribution of income is the lognormal distribution
\cite{Montroll}.  The lognormal distribution has the expression
\begin{equation}  \label{lognDistr}
  P_{LN}(r)=\frac{1}{r\sqrt{2\pi\sigma^2}}\exp\left[-
  \frac{(\ln r-\mu)^2}{2\sigma^2}\right], 
\end{equation}
where $\mu$ and $\sigma$ are two parameters.  As given by
(\ref{lognDistr}) the functional form of the lognormal distribution
makes it hard to compare it with data.  But the function
$\tilde{P}(r)=rP_{LN}(r)$ is a quadratic function in a log-log plot.
In Fig.\ \ref{fig:ln97} we plotted $\ln \tilde{P}(r)$ versus $\ln r$
for the 1997 income data from IRS \cite{Pub1304}.  As it can be seen
from Fig.\ \ref{fig:ln97} a fit of $\tilde{P}(r)$ with a quadratic
function cannot be made.  The data points show the power-law regime
for high incomes.  We conclude that the lognormal distribution does
not describe the distribution of income.
 
The distribution of wealth has essentially the same structure as the
distribution of income. An exponential regime for low wealth values
and a power-law tail for high wealth values.  Contrary to the
distribution of income, the distribution of wealth appears to have a
continuous derivative in the cross-over between the two regimes (see
the discussion at the beginning of Sec.\ \ref{sec:pltbc}).  The
shape of the wealth distribution suggests a modification of the
regular exponential Boltzmann-Gibbs distribution for values of wealth
$w\gg W$.  In the past years, in the context of non-extensive
statistical mechanics, Tsallis has proposed a viable alternative to
the Boltzmann-Gibbs distribution \cite{TsallisBook}.

The Tsallis distribution can be obtained by maximizing a generalized
entropy and is the subject of a lot of current research.  The appeal
of the Tsallis distribution is that it has power-law tails for large
arguments, and that in certain limit it becomes the exponential
distribution.  The Tsallis distribution \cite{TsallisBook} has the
expression
\begin{equation} \label{TsaDistr}
   P_{q}(w)=\frac{1}{W_T}\left[1+(q-1)\frac{w}{W_T}\right]^{\frac{q}{1-q}},
\end{equation}
where $W_T$ is the dimensional parameter of the Tsallis distribution, 
analogous to the temperature of the exponential distribution.  In the 
limit $q\to 1$, the distribution goes to the exponential form 
$P_{q=1}(w)\propto e^{-w/W_T}$. For $q>1$, the distribution has 
a power-law decay $P_{q}(w)\sim (1/w)^{q/q-1}$ for values $w\gg W$.  

\begin{figure}
\centerline{
\epsfig{file=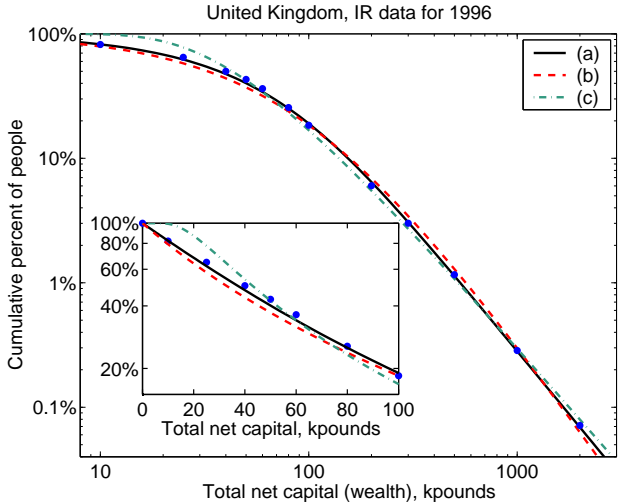,width=0.95\linewidth}}
\caption{ {\small
  Points: Internal Revenue wealth data for individuals, 1996
  \cite{UKwealth}.  Curve (a): Fit with the Kaniadakis distribution
  (\ref{KaniDistr}).  Curve (b): Fit with the Tsallis distribution
  (\ref{TsaDistr}).  Curve (c): Fit with the
  Bouchaud-Mezard/Solomon-Richmond distribution (\ref{BMS}).  The
  inset shows the same three curves in a log-linear scale.  The
  deviations of model (\ref{BMS}) from data are evident for small
  values of income.}}
\label{fig:tsallis96IR}
\end{figure}

In Fig.\ \ref{fig:tsallis96IR} we present the fit of UK wealth data
with the Tsallis distribution.  Overall, the quality of the fit is
good.  The parameters of the Tsallis distribution implied by the 
fit are: $q=1.42$, and for the Tsallis temperature $W_T=41 k\pounds$. 

Another distribution proposed by Kaniadakis \cite{Kaniadakis} in the
context of non-extensive statistical mechanics has the form of a
deformed exponential
\begin{equation} \label{KaniDistr}
  P_{\kappa}(x)=\left(\sqrt{1+\kappa^2 x^2} - 
  \kappa x\right)^{1/\kappa},
\end{equation}
where $x=w/W_\kappa$.  For $\kappa\to 0$, the distribution has the
exponential form $P_{\kappa=0}(w)=e^{-w/W_\kappa}$.  For large arguments
$w\gg W$, the distribution (\ref{KaniDistr}) has the power-law
behavior $P_{\kappa}(w)\approx (1/w)^{1/\kappa}$.  The fit with the 
Kaniadakis distribution is even better than that with the Tsallis 
distribution, as shown in Fig.\ \ref{fig:tsallis96IR}.  The parameters 
of the fit for the Kaniadakis distribution are: $\kappa=0.33$ and 
for the Kaniadakis temperature $W_\kappa=45$k$\pounds$.

A distribution for wealth put forward by Bouchaud and M\'ezard
\cite{Bouchaud}, and independently by Solomon and Richmond
\cite{SolomonRichmond}, is
\begin{equation} \label{BMS}
  P_{\alpha}(x)=\frac{(\alpha-1)^\alpha}{\Gamma(\alpha)}
  \frac{\exp(-\frac{\alpha-1}{x})}{x^{\alpha+1}},  
\end{equation}
where $x=w/W_c$. This distribution has an exponentially sharp cut-off
for wealths $w\ll W_c$, and a power-law tail $P_\alpha(w)\propto
1/w^{\alpha+1}$ for $w\gg W_c$. The fit of the UK wealth data with the
distribution (\ref{BMS}) is presented in Fig.\ \ref{fig:tsallis96IR}.
The values we obtain for the parameters are: $\alpha=1.93$ and
$W_c=74$ k$\pounds$.  The distribution (\ref{BMS}) has its maximum at
$x_{max}=(\alpha-1)/(\alpha+1)=0.32$, which essentially states that
for $w>0.32W_c=23$k$\pounds$ the distribution is a power law.  As seen
in Fig.\ \ref{fig:tsallis96IR} this claim is not supported by data.
Moreover, the percent of people in the power-law tail predicted by the
distribution (\ref{BMS}), $\int_{x_{max}}^\infty dx P_\alpha(x)=81\%$,
is clearly much larger than what is observed from the Lorenz curve for
wealth, Fig.\ \ref{fig:UKwealth_right}. 

All three proposed distributions
(\ref{TsaDistr},\ref{KaniDistr},\ref{BMS}) seem to capture well the
power-law tail.  In the inset of Fig.\ \ref{fig:tsallis96IR} a
log-linear scale is used, and this makes the discrepancy between
(\ref{BMS}) and data much more evident.  The distribution (\ref{BMS})
is clearly inappropriate for low incomes.

\subsection{Conclusions}
\label{sec:conc2}

Our analysis of the data shows that there are two clear regimes in the
distribution of individual income.  For low and moderate incomes up
to approximately 95\% of the total population, the distribution is
well described by an exponential, while the income of the top 5\%
individuals is described by a power-law (Pareto) regime. 

The exponential Boltzmann-Gibbs distribution naturally applies to
quantities that obey a conservation law, such as energy or money
\cite{DY-income}.  However, there is no fundamental reason why the sum of
incomes (unlike the sum of money) must be conserved.  Indeed, income
is a term in the time derivative of one's money balance (the other
term is spending).  Maybe incomes obey an approximate conservation
law, or somehow the distribution of income is simply proportional to
the distribution of money, which is exponential \cite{DY-income}.

Another explanation involves hierarchy.  Groups of people have
leaders, which have leaders of a higher order, and so on.  The number
of people decreases geometrically (exponentially) with the
hierarchical level.  If individual income increases linearly with the
hierarchical level, then the income distribution is exponential.
However, if income increases multiplicatively, then the distribution
follows a power law \cite{hierarchy}.  For moderate incomes below
\$100,000, the linear increase may be more realistic.  A similar
scenario is the Bernoulli trials \cite{Feller}, where individuals have
a constant probability of increasing their income by a fixed amount.

We found scaling in the cumulative probability distributions $N(r)$ of
individual income $r$ derived from the tax statistics for different
years in the UK and for different states in the US.  The distributions
$N_i(r)$ have the scaling form $f(r/R_i)$, where the scale $R_i$ (the
temperature) varies from one data set $i$ to another, but the scaling
function $f$ does not.  The function $f$ has an exponential
(Boltzmann-Gibbs) form at the low end, which covers about 95\% of
individuals.  At the high end, it follows a power (Pareto) law with
the exponents about 2.1 for the UK and 1.7 for the US.  Wealth
distribution in the UK also has a qualitatively similar shape with the
exponent about 1.9 and the temperature $W_{UK}=60$~k$\pounds$.  Some
of the other values of the exponents found in literature are 1.5
proposed by Pareto himself ($\alpha=1.5$), 1.36 found by Levy and
Solomon \cite{Solomon97} for the distribution of wealth in the Forbes
400 list, and 2.05 found by Souma \cite{Souma} for the high end of
income distribution in Japan.  The latter study is similar to our work
in the sense that it also uses tax statistics and explores the whole
range of incomes, not just the high end.  Souma \cite{Souma} finds
that the probability density $P(r)$ at lower incomes follows a
log-normal law with a maximum at a nonzero income.  This is in
contrast to our results, which suggest that the maximum of $P(r)$ is
at zero income.  The disrepancy may be due to the high threshold for
tax reporting in Japan, which distorts the data at the low end.  On
the other hand, if the data is indeed valid, it may reflect the actual
difference between the social stuctures of the US/UK and Japan.

\section{Distribution of stock-price fluctuations} 
\label{ch:3}
 
\subsection{Introduction}

Stochastic dynamics of stock prices is commonly described by a
geometric (multiplicative) Brownian motion, which gives a log-normal
probability distribution for stock price changes (returns)
\cite{Wilmott}.  However, numerous observations show that the tails of
the distribution decay slower than the log-normal distribution
predicts (the so-called ``fat-tails'' effect)
\cite{Bouchaud-book,Stanley-book,Voit}.  Particularly, much attention
was devoted to the power-law tails \cite{Mandelbrot2,Stanley-returns}.
The geometric Brownian motion model has two parameters: the drift
$\mu$, which characterizes the average growth rate, and the volatility
$\sigma$, which characterizes the noisiness of the process.  There is
empirical evidence and a set of stylized facts indicating that
volatility, instead of being a constant parameter, is driven by a
mean-reverting stochastic process \cite{Engle,Papanicolau}.  Various
mathematical models with stochastic volatility have been discussed in
literature \cite{Group,SteinStein,Heston,Baaquie}.

In this chapter, we study a particular stochastic volatility model,
where the square root of the stock-price volatility, called the
variance, follows a random process known in financial literature as
the Cox-Ingersoll-Ross process and in mathematical statistics as the
Feller process \cite{Papanicolau,Heston}.  We solve the Fokker-Planck
equation for this model exactly and find the joint probability
distribution of returns and variance as a function of time,
conditional on the initial value of variance.  The solution is
obtained in two different ways: using the method of characteristics
\cite{CourantHilbert} and the method of path integrals
\cite{Feynman,Schulman,Kleinert}.  The latter is more familiar to
physicists working in finance \cite{Baaquie,PIfinance}.

While returns are readily known from a financial time series data,
variance is not given directly, so it acts as a hidden stochastic
variable.  Thus, we integrate the joint probability distribution over
variance and obtain the mar\-gi\-nal probability distribution of returns
\textit{un}conditional on variance.  The latter distribution can be
directly compared with financial data.  We find excellent agreement
between our results and the Dow-Jones data for the period of
1982--2001.  Using only four fitting parameters, our equations very
well reproduce the probability distribution of returns for time lags
between 1 and 250 trading days.  This is in contrast to popular ARCH,
GARCH, EGARCH, TARCH, and similar models, where the number of
parameters can easily go to a few dozen \cite{McMillan}.

Our result for the probability distribution of returns has the form of
a one-dimensional Fourier integral, which is easily calculated
numerically or, in certain asymptotical limits, analytically.  For
large returns, we find that the probability distribution is
exponential in log-returns, which implies a power-law distribution for
returns, and we calculate the time dependence of the corresponding
exponents.  In the limit of long times, the probability distribution
exhibits scaling.  It becomes a function of a single combination of
return and time, with the scaling function expressed in terms of a
Bessel function.  The Dow-Jones data follow the predicted scaling
function for seven orders of magnitude.

The original theory of option pricing was developed by Black and
Scholes for the geometric Brownian motion model \cite{Wilmott}.
Numerous attempts to improve it using stochastic volatility models
have been made \cite{Papanicolau,Group,SteinStein,Heston,Baaquie}.
Particularly, option pricing for the same model as in our paper
\cite{DY-stock} was investigated by Heston \cite{Heston}.  Empirical
studies \cite{options} show that Heston's theory fares better than the
Black-Scholes model, but still does not fully capture the real-market
option prices.  Since our paper \cite{DY-stock} gives a closed-form
time-dependent expression for the probability distribution of returns
that agrees with financial data, it can serve as a starting point for
a better theory of option pricing.

\subsection{The model}
\label{sec:model}

We consider a stock, whose price $S_t$, as a function of time $t$,
obeys the stochastic differential equation of a geometric
(multiplicative) Brownian motion \cite{Wilmott}:
\begin{equation} \label{eqS}
  dS_t = \mu S_t\, dt + \sigma_t S_t\, dW_t^{(1)}.
\end{equation}
Here the subscript $t$ indicates time dependence, $\mu$ is the drift
parameter, $W_t^{(1)}$ is the standard random Wiener
process\footnote{The infinitesimal increments of the Wiener process
$dW_t$ are normally distributed (Gaussian) random variables with zero
mean and the variance equal to $dt$.}, and $\sigma_t$ is a
time-dependent parameter, called the stock volatility, which
characterizes the noisiness of the Wiener process.

Since any solution of (\ref{eqS}) depends only on $\sigma_t^2$, it is
convenient to introduce the new variable $v_t=\sigma_t^2$, which is
called the variance.  We assume that $v_t$ obeys the following
mean-reverting stochastic differential equation:
\begin{equation} \label{eqVar}
  dv_t = -\gamma(v_t - \theta)\,dt + \kappa\sqrt{v_t}\,dW_t^{(2)}.
\end{equation}
Here $\theta$ is the long-time mean of $v$, $\gamma$ is the rate of
relaxation to this mean, $W_t^{(2)}$ is the standard Wiener process,
and $\kappa$ is a parameter that we call the variance noise.  Eq.\
(\ref{eqVar}) is known in financial literature as the
Cox-Ingersoll-Ross (CIR) process and in mathematical statistics as the
Feller process \cite[p.\ 42]{Papanicolau}.  Alternative equations for
$v_t$, with the last term in (\ref{eqVar}) replaced by
$\kappa\,dW_t^{(2)}$ or by $\kappa v_t\,dW_t^{(2)}$, are also
discussed in literature \cite{Group}.  However, in this chapter, we
study only the case given by Eq.\ (\ref{eqVar}).

We take the Wiener process appearing in (\ref{eqVar}) to be correlated
with the Wiener process in (\ref{eqS}):
\begin{equation} \label{rho}
   dW_t^{(2)} = \rho\,dW_t^{(1)} + \sqrt{1-\rho^2}\,dZ_t, 
\end{equation} 
where $Z_t$ is a Wiener process independent of $W_t^{(1)}$, and
$\rho\in[-1,1]$ is the correlation coefficient.  A negative
correlation ($\rho<0$) between $W_t^{(1)}$ and $W_t^{(2)}$ is known as
the leverage effect \cite[p.\ 41]{Papanicolau}.

It is convenient to change the variable in (\ref{eqS}) from price
$S_t$ to log-return $r_t=\ln(S_t/S_0)$.  Using It\^{o}'s formula
\cite{Gardiner}, we obtain the equation satisfied by $r_t$:
\begin{equation}\label{eqR}
  dr_t = \left(\mu - \frac{v_t}{2}\right)dt +
  \sqrt{v_t}\,dW_t^{(1)}.
\end{equation}
The parameter $\mu$ can be eliminated from (\ref{eqR}) by changing the
variable to $x_t=r_t-\mu t$, which measures log-returns relative to
the growth rate $\mu$:
\begin{equation}\label{eqX}
   dx_t = - \frac{v_t}{2}\,dt + \sqrt{v_t}\,dW_t^{(1)}.
\end{equation}
Where it does not cause confusion with $r_t$, we use the term
``log-return'' also for the variable $x_t$.

Equations (\ref{eqX}) and (\ref{eqVar}) define a two-dimensional
stochastic process for the variables $x_t$ and $v_t$.  This process is
characterized by the transition probability $P_t(x,v\,|\,v_i)$ to have
log-return $x$ and variance $v$ at time $t$ given the initial
log-return $x=0$ and variance $v_i$ at $t=0$.  Time evolution of
$P_t(x,v\,|\,v_i)$ is governed by the Fokker-Planck (or forward
Kolmogorov) equation \cite{Gardiner}
\begin{eqnarray}\label{FP}
  && \frac{\partial}{\partial t}P = 
     \gamma\frac{\partial}{\partial v}\left[(v-\theta)P\right]
     + \frac12\frac{\partial}{\partial x}(vP)
\\ 
  && +\rho\kappa\frac{\partial^2}{\partial x\,\partial v}(vP)
     +\frac12\frac{\partial^2}{\partial x^2}(vP)  
     +\frac{\kappa^2}{2}\frac{\partial^2}{\partial v^2}(vP).  
\nonumber
\end{eqnarray}
The initial condition for (\ref{FP}) is a product of two delta
functions
\begin{equation} \label{initial}
        P_{t=0}(x,v\,|\,v_i)=\delta(x)\,\delta(v-v_i).
\end{equation}

The variance $v$ is a positive quantity, so Eq.\ (\ref{FP}) is defined
only for $v>0$.  However, when solving (\ref{FP}), it is convenient to
extend the domain of $v$ so that $v\in(-\infty,\infty)$.  If
$2\gamma\theta>\kappa^2$, such an extension does not change the
solution of the equation, because, given that $P=0$ for $v<0$ at the
initial time $t=0$, the condition $P_t(x,v<0)=0$ is preserved for all
later times $t>0$.  In order to demonstrate this, let us consider
(\ref{FP}) in the limit $v\to0$:
\begin{equation}\label{v->0}
  \frac{\partial}{\partial t}P = 
  -\left(\gamma\theta-\frac{\kappa^2}{2}\right)\frac{\partial P}{\partial v}
  +\gamma P + \rho\kappa\frac{\partial P}{\partial x}.
\end{equation}
Eq.\ (\ref{v->0}) is a first-order partial differential equation
(PDE), which describes propagation of $P$ from negative $v$ to
positive $v$ with the positive velocity $\gamma\theta-\kappa^2/2$.
Thus, the nonzero function $P(x,v>0)$ does not propagate to $v<0$, and
$P_t(x,v<0)$ remains zero at all times $t$.  Alternatively, it is
possible to show \cite[p.\ 67]{Wilmott} that, if
$2\gamma\theta>\kappa^2$, the random process (\ref{eqVar}) starting in
the domain $v>0$ can never reach the domain $v<0$.

The probability \hfill distribution \hfill of \hfill the \hfill variance 
\hfill itself, $\Pi_t(v)=\int dx\,P_t(x,v)$, satisfies the equation
\begin{equation} \label{A1varFP}
  \frac{\partial}{\partial t}\Pi_t(v) = 
  \frac{\partial}{\partial v}\left[\gamma(v-\theta)\Pi_t(v)\right]
  +\frac{\kappa^2}{2}\frac{\partial^2}{\partial^2 v}\left[v\Pi_t(v)\right],
\end{equation}
which is obtained from (\ref{FP}) by integration over $x$.  Eq.\
(\ref{A1varFP}) has the stationary solution
\begin{equation} \label{Pi_s}
   \Pi_\ast(v) = \frac{\alpha^{\beta+1}}{\Gamma(\beta+1)} v^\beta
   e^{-\alpha v}, \quad \alpha=\frac{2\gamma}{\kappa^2},
   \quad \beta=\alpha\theta-1,
\end{equation}
which is the Gamma distribution.  The maximum of $\Pi_\ast(v)$ is
reached at $v_{max}=\beta/\alpha=\theta-\kappa^2/2\gamma$.  The
width $w$ of $\Pi_\ast(v)$ can be estimated using the curvature at the
maximum $w\approx(\kappa^2/2\gamma)\sqrt{2\gamma\theta/\kappa^2-1}$.
The shape of $\Pi_\ast(v)$ is characterized by the dimensionless ratio
\begin{equation} \label{width}
   \chi = \frac{v_{max}}{w} = 
   \sqrt{\frac{2\gamma\theta}{\kappa^2} - 1}.
\end{equation}
When $2\gamma\theta/\kappa^2\to\infty$, $\chi\to\infty$ and
$\Pi_\ast(v)\to\delta(v-\theta)$.

\subsection{Solution of the Fokker-Planck equation}

Since $x$ appears in (\ref{FP}) only in the derivative operator
$\partial/\partial x$, it is convenient to make the Fourier transform
\begin{equation}\label{FT}
   P_t(x,v\,|\,v_i)=\frac{1}{2\pi}
   \int_{-\infty}^{+\infty}\!\! dp_x \, 
   e^{ip_x x} \overline{P}_{t,p_x}(v\,|\,v_i).
\end{equation}
Inserting (\ref{FT}) into (\ref{FP}), we find
\begin{eqnarray} \label{FourSch}
   && \frac{\partial}{\partial t}\overline{P} = 
      \gamma\frac{\partial}{\partial v}
      \left[(v-\theta)\overline{P}\right]
\\
   && -\left[\frac{p_x^2 -ip_x}{2} v
   - i\rho\kappa p_x\frac{\partial}{\partial v} v 
   - \frac{\kappa^2}{2}\frac{\partial^2}{\partial v^2}
   v\right]\overline{P}.  \nonumber 
\end{eqnarray}
Eq.\ (\ref{FourSch}) is simpler than (\ref{FP}), because the number of
variables has been reduced to two, $v$ and $t$, whereas $p_x$ only
plays the role of a parameter.

Since Eq.~(\ref{FourSch}) is linear in $v$ and quadratic in
$\partial/\partial v$, it can be simplified by taking the Fourier
transform over $v$
\begin{equation} \label{FourPV}
   \overline{P}_{t,p_x}(v\,|\,v_i) = \frac{1}{2\pi}
   \int_{-\infty}^{+\infty}\!\! dp_v \, 
   e^{ip_v v}\widetilde{P}_{t,p_x}(p_v\,|\,v_i).
\end{equation}
The PDE satisfied by $\widetilde{P}_{t,p_x}(p_v\,|\,v_i)$ is of the
first order
\begin{equation}\label{FourFour}
   \left[ \frac{\partial}{\partial t} + \left(
   \Gamma p_v+\frac{i\kappa^2}{2}p_v^2 +\frac{ip_x^2+p_x}{2}
   \right)\frac{\partial}{\partial p_v}\right] \widetilde{P}
   = -i\gamma\theta p_v \widetilde{P},
\end{equation}
where we introduced the notation 
\begin{equation} \label{Gamma}
    \Gamma = \gamma + i\rho\kappa p_x.
\end{equation}
Eq.~(\ref{FourFour}) has to be solved with the initial condition
\begin{equation} \label{v_i}
  \widetilde{P}_{t=0,p_x}(p_v\,|\,v_i)=\exp(-ip_v v_i).
\end{equation}

The solution of the PDE (\ref{FourFour}) is given by the method of
characteristics \cite{CourantHilbert}:
\begin{equation} \label{Kpv}
   \widetilde{P}_{t,p_x}(p_v\,|\,v_i) = \exp\left(-i\tilde{p}_v(0)v_i
   -i\gamma\theta\int_{0}^{t} d\tau\,\tilde{p}_v(\tau)\right),
\end{equation}
where the function $\tilde{p}_v(\tau)$ is the solution of the
characteristic (ordinary) differential equation
\begin{equation}\label{charEq}
   \frac{d\tilde{p}_v(\tau)}{d\tau} = \Gamma \tilde{p}_v(\tau) 
   +\frac{i\kappa^2}{2}\tilde{p}_v^2(\tau)  
   +\frac{i}{2}(p_x^2 - ip_x)
\end{equation}
with the boundary condition $\tilde{p}_v(t)=p_v$ specified at
$\tau=t$.  The solution (\ref{Kpv}) can be also obtained using the
method of path integrals described in Sec.\ \ref{path-integral}.
The differential equation (\ref{charEq}) is of the Riccati type with
constant coefficients \cite{Bender}, and its solution is
\begin{equation} \label{pvSol}
  \tilde{p}_v(\tau) =  
  -i\frac{2\Omega}{k^2}\frac{1}{\zeta e^{\Omega(t-\tau)} - 1} 
  + i\frac{\Gamma-\Omega}{k^2},
\end{equation}
where we introduced the frequency
\begin{equation} \label{eqOmega}
   \Omega=\sqrt{\Gamma^2 + \kappa^2(p_x^2-ip_x)}.  
\end{equation}
and the coefficient 
\begin{equation} \label{zeta}
   \zeta = 1 - i\frac{2\Omega}{\kappa^2 p_v - i(\Gamma-\Omega)}.
\end{equation}

Substituting (\ref{pvSol}) into (\ref{Kpv}) and taking the Fourier
transforms (\ref{FT}) and (\ref{FourPV}), we get the solution
\begin{eqnarray} \label{solution}
   && P_t(x,v\,|\,v_i)=\frac{1}{(2\pi)^2}
   \int\!\!\int_{-\infty}^{+\infty}\!\! dp_x\,dp_v \, 
   e^{ip_x x + ip_v v}
\\
   && \times\exp\left\{- i\tilde{p}_v(0)v_i
   + \frac{\gamma\theta(\Gamma-\Omega)t}{\kappa^2}
   - \frac{2\gamma\theta}{\kappa^2}\ln
   \frac{\zeta - e^{-\Omega t}}{\zeta - 1}\right\}
\nonumber
\end{eqnarray}
of the original Fokker-Planck equation (\ref{FP}) with the initial
condition (\ref{initial}), where $\tilde p_v(\tau=0)$ is given by
(\ref{pvSol}).

\subsection{Path-Integral Solution}
\label{path-integral}

The Fokker-Planck equation (\ref{FourSch}) can be though of as a
Schr\"{o}dinger equation in imaginary (Euclidean) time:
\begin{equation}
   \frac{\partial}{\partial t}\overline{P}_{t,p_x}(v\,|\,v_i) = -
   \hat{H}_{p_x}(\hat{p}_v,\hat{v})\overline{P}_{t,p_x}(v\,|\,v_i)
\end{equation}
with the Hamiltonian
\begin{equation} \label{H}
   \hat{H}=
   \frac{\kappa^2}{2}\hat{p}_v^2\hat{v} 
   - i\gamma\hat{p}_v(\hat{v}-\theta) 
   + \frac{p_x^2 -ip_x}{2}\hat{v}
   + \rho\kappa p_x\hat{p}_v\hat{v}.
\end{equation}
In (\ref{H}) we treat $\hat{p}_v$ and $\hat{v}$ as canonically
conjugated operators with the commutation relation
$[\hat{v},\hat{p}_v]=i$.  The transition probability
$\overline{P}_{p_x}(v,t\,|\,v_i)$ is the matrix element of the
evolution operator $\exp(-\hat{H}t)$ and has a path-integral
representation \cite{Feynman,Schulman,Kleinert}
\begin{equation}\label{propVar}
  \overline{P}_{t,p_x}(v\,|\,v_i) =
  \langle v|e^{-\hat{H}t}|v_i\rangle =
  \int\!\!{\cal{D}}v\,{\cal{D}}p_v\,e^{S_{p_x}[p_v(\tau),v(\tau)]}.
\end{equation}
Here the action functional $S_{p_x}[p_v(\tau),v(\tau)]$ is
\begin{equation}\label{action}
   S_{p_x} = \int_{0}^{t}\!\! d\tau\left\{ ip_v(\tau)\dot{v}(\tau) 
    - H_{p_x}\left[p_v(\tau),v(\tau)\right]\right\},
\end{equation}
and the dot denotes the time derivative.  The phase-space path
integral (\ref{propVar}) is the sum over all possible paths
$p_v(\tau)$ and $v(\tau)$ with the boundary conditions $v(\tau=0)=v_i$
and $v(\tau=t)=v$ imposed on $v$.

It is convenient to integrate the first term on the r.h.s.\ of
(\ref{action}) by parts:
\begin{eqnarray}
   && S_{p_x} = i[p_v(t)v-\tilde p_v(0)v_i] -
   i\gamma\theta\int_{0}^{t}\!\! d\tau\,p_v(\tau) \nonumber \\ && -
   \int_{0}^{t}\!\! d\tau \left[i\dot{p}_v(\tau) + \frac{\delta
   H}{\delta v(\tau)}\right]v(\tau),
\label{action'}
\end{eqnarray}
where we also separated the terms linear in $v(\tau)$ from the
Hamiltonian (\ref{H}).  Because $v(\tau)$ enters linearly in the
action (\ref{action'}), taking the path integral over ${\cal{D}}v$
generates the delta-functional
$\delta\left[p_v(\tau)-\tilde{p}_v(\tau)\right]$, where
$\tilde{p}_v(\tau)$ is the solution of the ordinary differential
equation (\ref{charEq}) with a boundary condition specified at
$\tau=t$.  Taking the path integral over ${\cal{D}}p_v$ resolves the
delta-functional, and we find
\begin{equation} \label{path}
   \overline{P}_{t,p_x}(v\,|\,v_i) = 
   \int\limits_{-\infty}^{+\infty}\!\frac{dp_v}{2\pi}
   e^{i[p_vv-\tilde p_v(0)v_i]
   -i\gamma\theta\int_{0}^{t} d\tau\,\tilde{p}_v(\tau)},
\end{equation}
where $\tilde p_v(\tau=t)=p_v$.  Eq.\ (\ref{path}) coincides with
(\ref{Kpv}) after the Fourier transform (\ref{FourPV}).

\subsection{Averaging over variance}

Normally we are interested only in log-returns $x$ and do not care
about variance $v$.  Moreover, whereas log-returns are directly known
from financial data, variance is a hidden stochastic variable that has
to be estimated.  Inevitably, such an estimation is done with some
degree of uncertainty, which precludes a clear-cut direct comparison
between $P_t(x,v\,|\,v_i)$ and financial data.  Thus, we introduce the
probability distribution
\begin{equation} \label{P}
   P_t(x\,|\,v_i)=\int_{-\infty}^{+\infty}\!\!dv\,P_t(x,v\,|\,v_i),
\end{equation}
where the hidden variable $v$ is integrated out.  The integration of
(\ref{solution}) over $v$ generates the delta-function $\delta(p_v)$,
which effectively sets $p_v=0$.  Substituting the coefficient $\zeta$
from (\ref{zeta}) with $p_v=0$ into (\ref{solution}), we find
\begin{eqnarray} \label{finalex}
   && P_t(x\,|\,v_i) = \frac{1}{2\pi}\int_{-\infty}^{+\infty} 
   \!\!dp_x\, e^{i p_x x 
   - v_i\frac{p_x^2 - ip_x}{\Gamma + \Omega\coth{(\Omega t/2)}} }
\nonumber \\
   && \times\, e^{- \frac{2\gamma\theta}{\kappa^2}\ln\left(
   \cosh\frac{\Omega t}{2} +\frac{\gamma}{\Omega}\sinh\frac{\Omega t}{2}
   \right) + \frac{\gamma\Gamma \theta t}{\kappa^2}}.
\end{eqnarray}

To check the validity of (\ref{finalex}), let us consider the case
$\kappa=0$.  In this case, the stochastic term in (\ref{eqVar}) is
absent, so the time evolution of variance is deterministic:
\begin{equation} \label{detV}
  v(t)= \theta + (v_i - \theta)e^{-\gamma t}.
\end{equation}
Then process (\ref{eqX}) gives a Gaussian distribution for $x$,
\begin{equation}\label{K0}
   P_t^{(\kappa=0)}(x\,|\,v_i)=\frac{1}{\sqrt{2\pi t \overline{v}_t}}
   \exp\left(-\frac{(x+\overline{v}_t t/2)^2}{2\overline{v}_t t}\right),
\end{equation}
with the time-averaged variance $\overline{v}_t =
\frac{1}{t}\int_{0}^{t}d\tau\,v(\tau)$.  Eq.\ (\ref{K0}) demonstrates
that the probability distribution of stock price $S$ is log-normal in
the case $\kappa=0$.  On the other hand, by taking the limit
$\kappa\rightarrow0$ and integrating over $p_x$ in (\ref{finalex}), we
reproduce the same expression (\ref{K0}).

Eq.\ (\ref{finalex}) cannot be directly compared with financial time
series data, because it depends on the unknown initial variance $v_i$.
In order to resolve this problem, we assume that $v_i$ has the
stationary probability distribution $\Pi_\ast(v_i)$, which is given by
(\ref{Pi_s}).\footnote{An attempt to determine the probability
distribution of volatility empirically for the S\&P 500 stock index
was done in Ref.\ \cite{Stanley-volatility}.}  Thus we introduce the
probability distribution $P_t(x)$ by averaging (\ref{finalex}) over
$v_i$ with the weight $\Pi_\ast(v_i)$:
\begin{equation} \label{dv_i}
  P_t(x)= \int_0^\infty \!\!dv_i\,\Pi_\ast(v_i)\,P_t(x\,|\,v_i).
\end{equation}
The integral over $v_i$ is similar to the one of the Gamma function
and can be taken explicitly.  The final result is the Fourier integral
\begin{equation} \label{Pfinal}
   P_t(x) = \frac{1}{2\pi}\int_{-\infty}^{+\infty} \!\!dp_x\,
   e^{ip_x x + F_t(p_x)}
\end{equation}
with
\begin{eqnarray}
  && \!\!\!\!\!\!\!\!\!\!\!\! 
  F_t(p_x)=\frac{\gamma\Gamma\theta t}{\kappa^2} \nonumber \\
  && \!\!\!\!\!\!\!\!\!\!\!\!{} - \frac{2\gamma\theta}{\kappa^2}
  \ln\left[\cosh\frac{\Omega t}{2} + 
  \frac{\Omega^2 -\Gamma^2 + 2\gamma\Gamma}{2\gamma\Omega}
  \sinh\frac{\Omega t}{2}\right].
\label{phaseF}
\end{eqnarray}
The variable $p_x$ enters (\ref{phaseF}) via the variables $\Gamma$
from (\ref{Gamma}) and $\Omega$ from (\ref{eqOmega}).  It is easy to
check that $P_t(x)$ is real, because Re$F$ is an even function of
$p_x$ and Im$F$ is an odd one.  One can also check that
$F_t(p_x=0)=0$, which implies that $P_t(x)$ is correctly normalized at
all times: $\int dx\,P_t(x)=1$.  The second term in the r.h.s.\ of
(\ref{phaseF}) vanishes when $\rho=0$, i.e.\ when there are no
correlations between stock price and variance.  The simplified result
for the case $\rho=0$ is given in \ref{t>>1} by Eqs.\
(\ref{Pfinal'}), (\ref{phaseF'}), and (\ref{eqOmega'}).

\begin{figure}
\centerline{
\epsfig{file=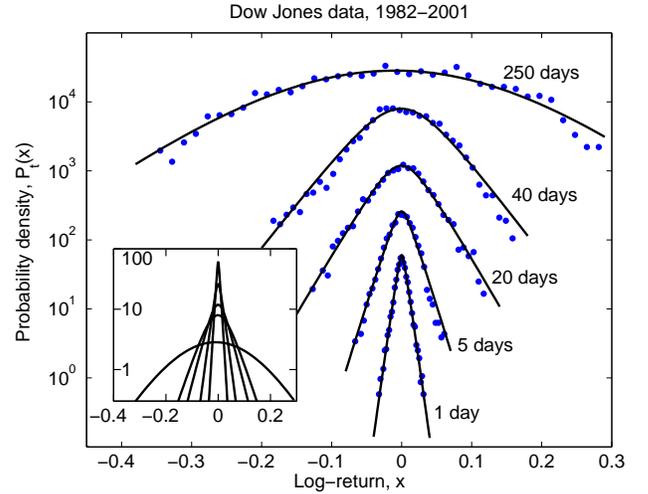,width=0.95\linewidth}}
\caption{{\small Probability distribution $P_t(x)$ of log-return $x$ for
  different time lags $t$.  Points: The Dow-Jones data for $t=1$, 5,
  20, 40, and 250 trading days.  Solid lines: Fit of the data with
  Eqs.\ (\ref{Pfinal}) and (\ref{phaseF}).  For clarity, the data
  points and the curves for successive $t$ are shifted up by the
  factor of 10 each. The inset shows the curves without vertical
  shift.}}
\label{fig:data}
\end{figure}

Eqs.\ (\ref{Pfinal}) and (\ref{phaseF}) for the probability
distribution $P_t(x)$ of log-return $x$ at time $t$ are the central
analytical result of the chapter.  The integral in (\ref{Pfinal}) can be
calculated numerically or, in certain regimes discussed in Secs.\
\ref{t>>1} and \ref{x>>1}, analytically.  In Fig.\ \ref{fig:data}, the
calculated function $P_t(x)$, shown by solid lines, is compared with
the Dow-Jones data, shown by dots.  (Technical details of the data
analysis are discussed in Sec.\ \ref{sec:data}.)  Fig.\ \ref{fig:data}
demonstrates that, with a fixed set of the parameters $\gamma$,
$\theta$, $\kappa$, $\mu$, and $\rho$, Eqs.\ (\ref{Pfinal}) and
(\ref{phaseF}) very well reproduce the distribution of log-returns $x$
of the Dow-Jones index for \textit{all} times $t$.  In the log-linear
scale of Fig.\ \ref{fig:data}, the tails of $\ln P_t(x)$ vs.\ $x$ are
straight lines, which means that that tails of $P_t(x)$ are
exponential in $x$.  For short times, the distribution is narrow, and
the slopes of the tails are nearly vertical.  As time progresses, the
distribution broadens and flattens.

\subsection{Asymptotic behavior for long time $t$}
\label{t>>1}

Eq.\ (\ref{eqVar}) implies that variance reverts to the equilibrium
value $\theta$ within the characteristic relaxation time $1/\gamma$.
In this section, we consider the asymptotic limit where time $t$ is
much longer than the relaxation time: $\gamma t\gg2$.  According to
(\ref{Gamma}) and (\ref{eqOmega}), this condition also implies that
$\Omega t\gg2$.  Then Eq.\ (\ref{phaseF}) reduces to
\begin{equation} \label{F_t>>1}
   F_t(p_x)\approx\frac{\gamma\theta t}{\kappa^2}(\Gamma-\Omega).
\end{equation}

Let us change of the variable of integration in (\ref{Pfinal}) to
\begin{equation} \label{tilde_p_x}
   p_x=\frac{\omega_0}{\kappa\sqrt{1-\rho^2}}\,\tilde p_x+ip_0,
\end{equation}
   where
\begin{equation} \label{p_0}
   p_0=\frac{\kappa-2\rho\gamma}{2\kappa(1-\rho^2)},\quad
   \omega_0=\sqrt{\gamma^2+\kappa^2(1-\rho^2)p_0^2}.
\end{equation}
Substituting (\ref{tilde_p_x}) into (\ref{Gamma}), (\ref{eqOmega}), and
(\ref{F_t>>1}), we transform (\ref{Pfinal}) to the following form
\begin{equation}\label{AppAll}
   P_t(x) = \frac{\omega_0e^{-p_0x+\Lambda t}}{\pi\kappa\sqrt{1-\rho^2}}
   \int_0^\infty\!\!d\tilde p_x\, \cos(A\tilde p_x) 
   e^{-B\sqrt{1+\tilde p_x^2}}, 
\end{equation}
where 
\begin{equation}
   A=\frac{\omega_0}{\kappa\sqrt{1-\rho^2}}\left(x 
   + \rho\frac{\gamma\theta t}{\kappa}\right),\quad
   B=\frac{\gamma\theta\omega_0 t}{\kappa^2},
\end{equation}
and
\begin{equation} \label{Lambda}
   \Lambda=\frac{\gamma\theta}{2\kappa^2}
   \frac{2\gamma-\rho\kappa}{1-\rho^2}.
\end{equation}
According \hfill to \cite{GR}, \hfill the \hfill integral \hfill in 
\hfill (\ref{AppAll}) \hfill is \hfill equal \hfill to \\ 
$B K_1(\sqrt{A^2+B^2})/\sqrt{A^2+B^2}$, 
where $K_1$ is the first-order modified Bessel function.  

\begin{figure}
\centerline{
\epsfig{file=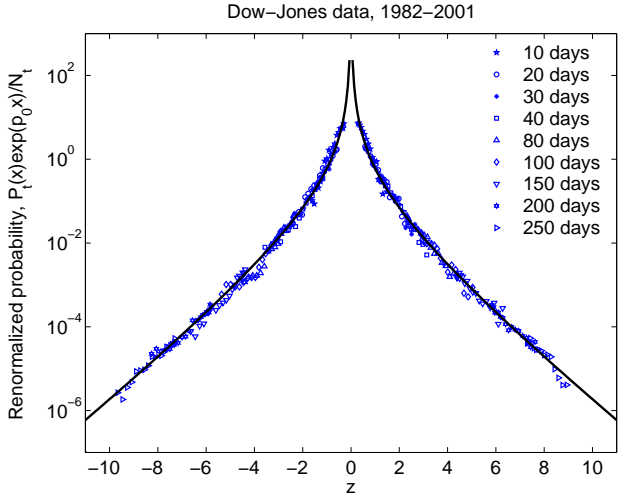,width=0.95\linewidth}}
\caption{{\small Renormalized probability density $P_t(x)e^{p_0x}/N_t$ plotted
   as a function of the scaling argument $z$ given by (\ref{K1arg}).
   Solid line: The scaling function $P_{\ast}(z)=K_1(z)/z$ from
   (\ref{Pbess}), where $K_1$ is the first-order modified Bessel
   function.  Symbols: The Dow-Jones data for different time lags
   $t$.}}
\label{fig:Bessel}
\end{figure}

Thus, Eq.\ (\ref{Pfinal}) in the limit $\gamma t\gg2$ can be
represented in the scaling form
\begin{equation} \label{Pbess}
   P_t(x)=N_t\,e^{-p_0x}P_{\ast}(z), \quad P_{\ast}(z)=K_1(z)/z,
\end{equation}
where the argument $z=\sqrt{A^2+B^2}$ is
\begin{equation} \label{K1arg}
   z=\frac{\omega_0}{\kappa}
   \sqrt{\frac{(x+\rho\gamma\theta t/\kappa)^2}{1-\rho^2} + 
   \left(\frac{\gamma\theta t}{\kappa}\right)^2},
\end{equation}
and the time-dependent normalization factor $N_t$ is
\begin{equation} \label{N}
   N_t=\frac{\omega_0^2\gamma\theta t}{\pi\kappa^3\sqrt{1-\rho^2}}
   \,e^{\Lambda t},
\end{equation}
Eq.\ (\ref{Pbess}) demonstrates that, up to the factors $N_t$ and
$e^{-p_0x}$, the dependence of $P_t(x)$ on the two arguments $x$ and
$t$ is given by the function $P_{\ast}(z)$ of the single scaling
argument $z$ in (\ref{K1arg}).  Thus, when plotted as a function of
$z$, the data for different $x$ and $t$ should collapse on the single
universal curve $P_{\ast}(z)$. This is beautifully illustrated by
Fig.\ \ref{fig:Bessel}, where the Dow-Jones data for different time
lags $t$ follows the curve $P_{\ast}(z)$ for seven decades.

In the limit $z\gg1$, we can use the asymptotic expression \cite{GR}
$K_1(z)\approx e^{-z}\sqrt{\pi/2z}$ in (\ref{Pbess}) and take the
logarithm of $P$.  Keeping only the leading term proportional to $z$
and omitting the subleading term proportional to $\ln z$, we find
\begin{equation} \label{P32}
   \ln\frac{P_t(x)}{N_t}\approx -p_0x-z \quad {\rm for} \quad z\gg1.
\end{equation}
Let us examine Eq.\ (\ref{P32}) for large and small $|x|$.

In the first case $|x|\gg\gamma\theta t/\kappa$, Eq.\ (\ref{K1arg})
gives $z\approx\omega_0|x|/\kappa\sqrt{1-\rho^2}$, so Eq.\ (\ref{P32})
becomes
\begin{equation} \label{PlargeX}
   \ln\frac{P_t(x)}{N_t}
   \approx-p_0x-\frac{\omega_0}{\kappa\sqrt{1-\rho^2}}|x|.
\end{equation}
Thus, the probability distribution $P_t(x)$ has the exponential tails
(\ref{PlargeX}) for large log-returns $|x|$.  Notice that, in the
considered limit $\gamma t\gg2$, the slopes $d\ln P/dx$ of the
exponential tails (\ref{PlargeX}) do not depend on time $t$.  Because
of $p_0$, the slopes (\ref{PlargeX}) for positive and negative $x$ are
not equal, thus the distribution $P_t(x)$ is not symmetric with
respect to up and down price changes.  According to (\ref{p_0}), this
asymmetry is enhanced by a negative correlation $\rho<0$ between stock
price and variance.

\begin{figure}
\centerline{
\epsfig{file=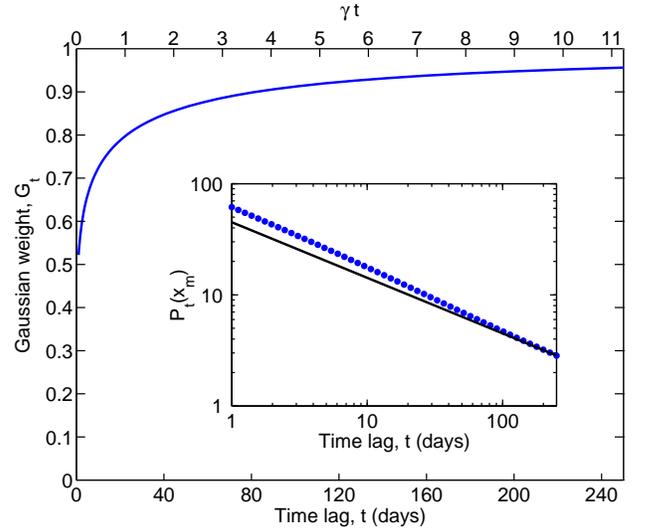,width=0.95\linewidth}}
\caption{{\small The fraction $G_t$ of the total probability contained in the
   Gaussian part of $P_t(x)$ vs.\ time lag $t$.  Inset: Time
   dependence of the probability density at maximum $P_t(x_m)$
   (points), compared with the Gaussian $t^{-1/2}$ behavior (solid
   line).}}
\label{fig:Gauss}
\end{figure}

In the second case $|x|\ll\gamma\theta t/\kappa$, by Taylor-expanding
$z$ in (\ref{K1arg}) near its minimum and substituting the result into
(\ref{P32}), we get
\begin{equation} \label{Gauss}
    \ln\frac{P_t(x)}{N'_t}\approx -p_0x -
    \frac{\omega_0(x+\rho\gamma\theta t/\kappa)^2}
    {2(1-\rho^2)\gamma\theta t},
\end{equation}
where $N'_t=N_t\exp(-\omega_0\gamma\theta t/\kappa^2)$.  Thus, for
small log-returns $|x|$, the probability distribution $P_t(x)$ is
Gaussian with the width increasing linearly in time.  The maximum of
$P_t(x)$ in (\ref{Gauss}) is achieved at
\begin{equation} \label{x_m}
   x_{m}(t)=-\frac{\gamma\theta t}{2\omega_0}\left(
   1+2\,\frac{\rho(\omega_0-\gamma)}{\kappa}\right).
\end{equation}
Eq.\ (\ref{x_m}) gives the most probable log-return $x_m(t)$ at time
$t$, and the coefficient in front of $t$ constitutes a correction to
the average growth rate $\mu$, so that the actual growth rate is
$\bar\mu=\mu+d x_m/dt$.

As Fig.\ \ref{fig:data} illustrates, $\ln P_t(x)$ is indeed linear in
$x$ for large $|x|$ and quadratic for small $|x|$, in agreement with
(\ref{PlargeX}) and (\ref{Gauss}).  As time progresses, the
distribution, which has the scaling form (\ref{Pbess}) and
(\ref{K1arg}), broadens.  Thus, the fraction $G_t$ of the total
probability contained in the parabolic (Gaussian) portion of the curve
increases, as illustrated in Fig.\ \ref{fig:Gauss}.  

In the remainder of the section, we explain the way the Gaussian
weight of the distribution has been calculated.  As explained in Sec.\
\ref{sec:data}, the case relevant for comparison with the data is
$\rho=0$.  In this case, by shifting the contour of integration in
(\ref{Pfinal}) as follows $p_x\to p_x+i/2$, we find
\begin{equation} \label{Pfinal'}
   P_t(x) = e^{-x/2}\int_{-\infty}^{+\infty} \frac{dp_x}{2\pi} \,
   e^{ip_x x + F_t(p_x)},
\end{equation}
where
\begin{equation} \label{phaseF'}
   F_t(p_x)=\frac{\gamma^2\theta t}{\kappa^2}  
   - \frac{2\gamma\theta}{\kappa^2}
   \ln\left[\cosh\frac{\Omega t}{2} + 
   \frac{\Omega^2+\gamma^2}{2\gamma\Omega}
   \sinh\frac{\Omega t}{2}\right]
\end{equation}
and
\begin{equation} \label{eqOmega'}
   \Omega=\sqrt{\gamma^2 + \kappa^2(p_x^2+1/4)}.
\end{equation}
Now the function $F_t(p_x)$ is real and symmetric:
$F_t(p_x)=F_t(-p_x)$.  Thus, the integral in (\ref{Pfinal'}) is a
symmetric function of $x$.  So it is clear that the only source of
asymmetry of $P_t(x)$ in $x$ is the exponential prefactor in
(\ref{Pfinal'}), as discussed at the end of Sec.\ \ref{sec:data}.
Eqs.\ (\ref{Pfinal'}), (\ref{phaseF'}), and (\ref{eqOmega'}) are much
simpler than those for $\rho\neq0$.

Let us expand the integral in (\ref{Pfinal'}) for small $x$:
\begin{equation} \label{expand}
   P_t(x)\approx e^{-x/2}\left(\mu_0-\frac12\mu_2 x^2\right),
\end{equation}
where the coefficients are the first and the second moments of
$\exp[F_t(p_x)]$
\begin{equation} \label{mu}
   \mu_0(t)=\int\limits_{-\infty}^{+\infty}\frac{dp_x}{2\pi}\,e^{F_t(p_x)},
   \quad 
   \mu_2(t)=\int\limits_{-\infty}^{+\infty}
   \frac{dp_x}{2\pi}\,p_x^2e^{F_t(p_x)}.
\end{equation}
On the other hand, we know that $P_t(x)$ is Gaussian for small $x$.
So we can write
\begin{equation} \label{approxGauss}
   P_t(x)\approx \mu_0\, e^{-x/2}e^{-\mu_2 x^2/2\mu_0},
\end{equation}
with the same coefficients as in (\ref{expand}).  If we ignore the
existence of fat tails and extrapolate (\ref{approxGauss}) to
$x\in(-\infty,\infty)$, the total probability contained in such a
Gaussian extrapolation will be
\begin{equation} \label{W}
   G_t=\int\limits_{-\infty}^{+\infty}dx\,\mu_0\,e^{-x/2-\mu_2
   x^2/2\mu_0}=\sqrt{\frac{2\pi\mu_0^3}{\mu_2}}\,
   e^{\mu_0/8\mu_2}.
\end{equation}
Obviously, $G_t<1$, because the integral (\ref{W}) does not take into
account the probability contained in the fat tails.  Thus, the
difference $1-G_t$ can be taken as a measure of how much the actual
distribution $P_t(x)$ deviates from a Gaussian function.

We calculate the moments (\ref{mu}) numerically for the function $F$
given by (\ref{phaseF'}), then determine the Gaussian weight $G_t$
from (\ref{W}) and plot it in Fig.\ \ref{fig:Gauss} as a function of
time.  For $t\to\infty$, $G_t\to1$, i.e.\ $P_t(x)$ becomes Gaussian
for very long time lags, which is known in literature
\cite{Stanley-returns}.  In the opposite limit $t\to0$, $F_t(p_x)$
becomes a very broad function of $p_x$, so we cannot calculate the
moments $\mu_0$ and $\mu_2$ numerically.  The singular limit $t\to0$
requires an analytical study.

Fig.\ \ref{fig:Gauss} shows that, at sufficiently long times, the total
probability contained in the non-Gaussian tails becomes negligible,
which is known in literature \cite{Stanley-returns}.  The inset in
Fig.\ \ref{fig:Gauss} illustrates that the time dependence of the
probability density at maximum, $P_t(x_m)$, is close to $t^{-1/2}$,
which is characteristic of a Gaussian evolution.

\subsection{Asymptotic behavior for large log-return $x$}
\label{x>>1}

\begin{figure}
\centerline{
\epsfig{file=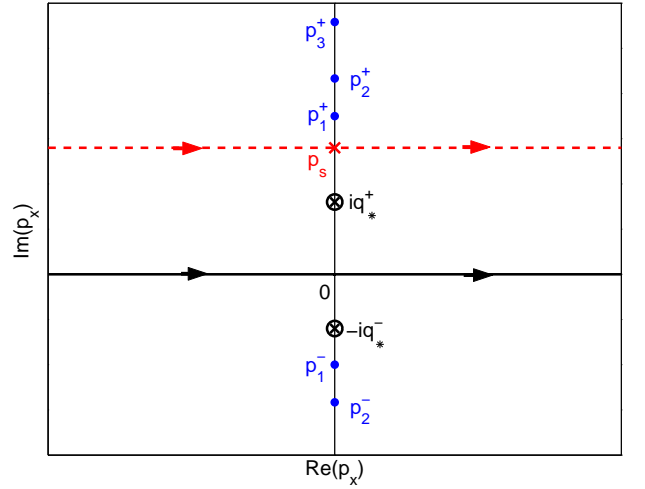,width=0.95\linewidth}}
\caption{{\small Complex plane of $p_x$. Dots: The singularities of
   $F_t(p_x)$. Circled crosses: The accumulation points $\pm iq_*^\pm$
   of the singularities in the limit $\gamma t\gg 2$.  Symbol
   $\times$: Saddle point $p_s$, which is located in the upper
   half-plane for $x>0$.  Dashed line: The contour of integration
   displaced from the real axis in order to pass through the saddle
   point $p_s$.}}
\label{fig:complex} 
\end{figure}

In the complex plane of $p_x$, function $F(p_x)$ becomes singular at
the points $p_x$ where the argument of any logarithm in (\ref{phaseF})
vanishes.  These points are located on the imaginary axis of $p_x$ and
are shown by dots in Fig.\ \ref{fig:complex}.  The singularity closest
to the real axis is located on the positive (negative) imaginary axis
at the point $p_1^+$ $(p_1^-)$.  At these two points, the argument of
the last logarithm in (\ref{phaseF}) vanishes, and we can approximate
$F(p_x)$ by the dominant, singular term:
$F(p_x)\approx-(2\gamma\theta/\kappa^2)\ln(p_x-p_1^\pm)$.

For large $|x|$, the integrand of (\ref{Pfinal}) oscillates very fast
as a function of $p_x$.  Thus, we can evaluate the integral using the
method of stationary phase \cite{Bender} by shifting the contour of
integration so that is passes through a saddle point of the argument
$ip_xx+F(p_x)$ of the exponent in (\ref{Pfinal}).  The saddle point
position $p_s$, shown in Fig.\ \ref{fig:complex} by the symbol
$\times$, is determined by the equation
\begin{equation} \label{saddle}
   ix=-\left.\frac{dF(p_x)}{dp_x}\right|_{p_x=p_s}
   \approx\frac{2\gamma\theta}{\kappa^2}\times
   \left\{\begin{array}{ll}
      \frac{1}{p_s-p_1^+},& x>0, \\
      \frac{1}{p_s-p_1^-},& x<0.
   \end{array}\right.
\end{equation}
For a large $|x|$ such that $|xp_1^\pm|\gg2\gamma\theta/\kappa^2$, the
saddle point $p_s$ is very close to the singularity point: $p_s\approx
p_1^+$ for $x>0$ and $p_s\approx p_1^-$ for $x<0$.  Then the
asymptotic expression for the probability distribution is
\begin{equation} \label{P_x>>1}
   P_t(x)\sim
   \left\{\begin{array}{ll}
      e^{-xq_t^+},& x>0, \\
      e^{ xq_t^-},& x<0,
   \end{array}\right.
\end{equation}
where $q_t^\pm=\mp ip_1^\pm(t)$ are real and positive.  Eq.\
(\ref{P_x>>1}) shows that, for all times $t$, the tails of the
probability distribution $P_t(x)$ for large $|x|$ are exponential.
The slopes of the exponential tails, $q^\pm=\mp\,d\ln P/dx$, are
determined by the positions $p_1^{\pm}$ of the singularities closest
to the real axis.

\begin{figure}
\centerline{
\epsfig{file=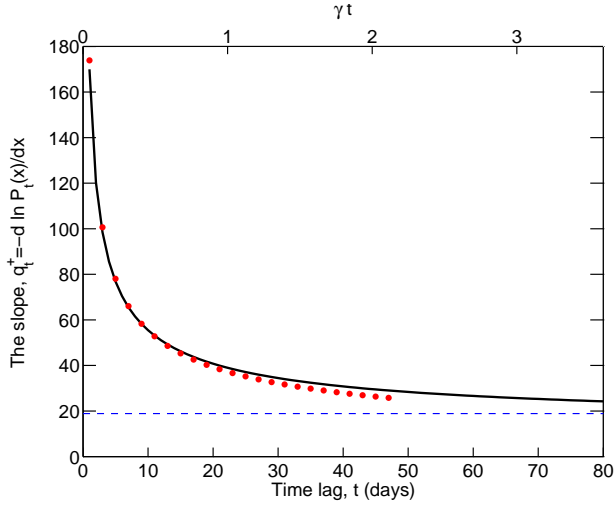,width=0.95\linewidth}}
\caption{{small Solid line: The slope $q_t^+=-d\ln P/dx$ of the exponential
   tail for $x>0$ as a function of time.  Points: The asymptotic
   approximation (\ref{p1pm}) for the slope in the limit $\gamma
   t\ll2$.  Dashed line: The saturation value $q_*^+$ for $\gamma
   t\gg2$, Eq.\ (\ref{q_*}).}}
\label{fig:slope} 
\end{figure}
 
These positions $p_1^{\pm}(t)$ and, thus, the slopes $q_t^\pm$ depend
on time $t$.  For times much shorter than the relaxation time ($\gamma
t\ll2$), the singularities lie far away from the real axis.  As time
increases, the singularities move along the imaginary axis toward the
real axis.  Finally, for times much longer than the relaxation time
($\gamma t\gg2$), the singularities approach limiting points:
$p_1^{\pm}\to\pm iq_*^\pm$, which are shown in Fig.\ \ref{fig:complex}
by circled $\times$'s.  Thus, as illustrated in Fig.\ \ref{fig:slope},
the slopes $q_t^\pm$ monotonously decrease in time and saturate at
long times:
\begin{equation} \label{q_*}
   q_t^\pm\to q_*^\pm= \pm p_0+\frac{\omega_0}{\kappa\sqrt{1-\rho^2}}
   \quad {\rm for} \quad \gamma t\gg2.
\end{equation}
The slopes (\ref{q_*}) are in agreement with Eq.\ (\ref{PlargeX})
valid for $\gamma t\gg2$.  The time dependence $q_t^{\pm}$ at short
times can be also found analytically:
\begin{equation} \label{p1pm}
   q_t^{\pm}\approx\pm
   p_0+\sqrt{p_0^2+\frac{4\gamma}{\kappa^2(1-\rho^2) t}} \quad {\rm for}
   \quad \gamma t\ll2.
\end{equation}
The dotted curve in Fig.\ \ref{fig:slope} shows that Eq.\ (\ref{p1pm})
works very well for short times $t$, where the slope diverges at
$t\to0$.

\subsection{Comparison with Dow-Jones time series}
\label{sec:data}

To test the model against financial data, we downloaded daily closing
values of the Dow-Jones industrial index for the period of 20 years
from 1 January 1982 to 31 December 2001 from the Web site of Yahoo
\cite{Yahoo}. The data set contains 5049 points, which form the time
series $\{S_\tau\}$, where the integer time variable $\tau$ is the
trading day number.  We do not filter the data for short days, such as
those before holidays.

\begin{figure}
\centerline{
\epsfig{file=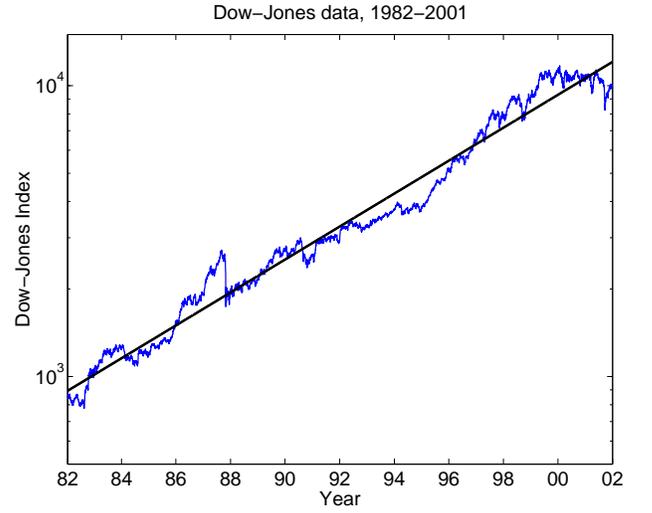,width=0.95\linewidth}}
\caption{{\small Log-linear plot of the Dow-Jones versus time, for the period
  1982-2001. The straight solid line through the data points
  represents a fit of the Dow-Jones index with an exponential function 
  with growth rate $\mu=13.3 \%/$year.}}
\label{fig:dowFit}
\end{figure}

\begin{figure}
\centerline{
\epsfig{file=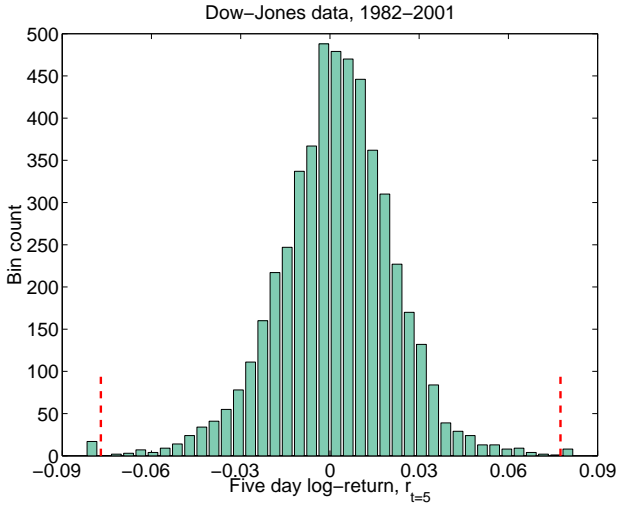,width=0.95\linewidth}}
\caption{{\small Histogram of five day log-returns $r_{t=5}$.  The first and 
  the last bin, which are separated from the rest by the two vertical
  dashed lines, are discarded in the fitting process.  The population
  of the first (last) bin contains all the extreme events from $-\infty
  (+\infty)$ to the left(right) dashed line.}}
\label{fig:prob-r5}
\end{figure}

Given $\{S_\tau\}$, we use the following procedure to extract the
probability density $P_t^{(DJ)}(r)$ of log-return $r$ for a given time
lag $t$.  For the fixed $t$, we calculate the set of log-returns
$\{r_\tau=\ln S_{\tau+t}/S_\tau\}$ for all possible times $\tau$.
Then we partition the $r$-axis into equally spaced bins of the width
$\Delta r$ and count the number of log-returns $r_\tau$ belonging to
each bin.  In this process, we omit the bins with occupation numbers
less than five, because we consider such a small statistics
unreliable.  Only less than 1\% of the entire data set is omitted in
this procedure.  Dividing the occupation number of each bin by $\Delta
r$ and by the total occupation number of all bins, we obtain the
probability density $P_t^{(DJ)}(r)$ for a given time lag $t$.  To find
$P_t^{(DJ)}(x)$, we replace $r\to x+\mu t$. In Fig.\ \ref{fig:dowFit}
we plot the value of the Dow-Jones index, and the fit with an
exponential function corresponding to an inflation rate $\mu$.  In
Fig.\ \ref{fig:prob-r5} we illustrate the histogram corresponding to
the probability density $P_t^{(DJ)}(r)$ for a time lag $t=5$ days.
The first and last bin in Fig.\ \ref{fig:prob-r5} show all the events
outside the two vertical dashed lines.  These events are very rare,
contribute less than $1\%$ to the entire number of data points, and
therefore they are discarded in the fitting process.

Assuming that the system is ergodic, so that ensemble averaging is
equivalent to time averaging, we compare $P_t^{(DJ)}(x)$ extracted
from the time series data and $P_t(x)$ calculated in previous
sections, which describes ensemble distribution.  In the language of
mathematical statistics, we compare our theoretically derived
population distribution with the sample distribution extracted from
the time series data.  We determine parameters of the model by
minimizing the mean-square deviation $\sum_{x,t}|\ln P_t^{(DJ)}(x)-\ln
P_t(x)|^2$, where the sum is taken over all available $x$ and $t=1$,
5, 20, 40, and 250 days.  These values of $t$ are selected because
they represent different regimes: $\gamma t\ll1$ for $t=1$ and 5 days,
$\gamma t\approx1$ for $t=20$ days, and $\gamma t\gg1$ for $t=40$ and
250 days.  As Figs.\ \ref{fig:data} and \ref{fig:Bessel} illustrate,
our expression (\ref{Pfinal}) and (\ref{phaseF}) for the probability
density $P_t(x)$ agrees with the data very well, not only for the
selected five values of time $t$, but for the whole time interval from
1 to 250 trading days.  However, we cannot extend this comparison to
$t$ longer than 250 days, which is approximately 1/20 of the entire
range of the data set, because we cannot reliably extract
$P_t^{(DJ)}(x)$ from the data when $t$ is too long.

The values obtained for the four fitting parameters ($\gamma$,
$\theta$, $\kappa$, $\mu$) are given in Table \ref{paramVal}.  Within
the scattering of the data, we do not find any discernible difference
between the fits with the correlation coefficient $\rho$ being zero or
slightly different from zero.  Thus, we conclude that the correlation
$\rho$ between the noise terms for stock price and variance in Eq.\
(\ref{rho}) is practically zero, if it exists at all.  Our conclusion
is in contrast with the value $\rho=-0.58$ found in \cite{Masoliver} by
fitting the leverage correlation function introduced in
\cite{Bouchaud-PRL}.  Further study is necessary in order to resolve
this discrepancy.  Nevertheless, all theoretical curves shown in this 
chapter are calculated for $\rho=0$, and they fit the data very well.

All four parameters ($\gamma$, $\theta$, $\kappa$, $\mu$) shown in
Table \ref{paramVal} have the dimensionality of 1/time.  The first
line of the Table gives their values in the units of 1/day, as
originally determined in our fit.  The second line shows the
annualized values of the parameters in the units of 1/year, where we
utilize the average number of 252.5 trading days per calendar year to
make the conversion.  The relaxation time of variance is equal to
$1/\gamma=22.2$ trading days = 4.4 weeks $\approx$ 1 month, where we
took into account that 1 week = 5 trading days.  Thus, we find that
variance has a rather long relaxation time, of the order of one month,
which is in agreement with the conclusion of Ref.\ \cite{Masoliver}.

Using the numbers given in Table \ref{paramVal}, we find the value of
the parameter $2\gamma\theta/\kappa^2=1.296$.  Since this parameter is
greater than one, the stochastic process (\ref{eqVar}) never reaches
the domain $v<0$, as discussed in Sec.\ \ref{sec:model}.

\begin{table} 
\begin{center} 
{\setlength{\tabcolsep}{3pt}
\begin{tabular}{|c|cccc|}
\hline
{\small Units} & {\small $\gamma$} & {\small $\theta$}  
     &  {\small $\kappa$} & {\small $\mu$} \\
\hline
{\small 1/day}  & {\small $4.50\times10^{-2}$}  & {\small$8.62\times10^{-5}$}  
     &  {\small $2.45\times10^{-3}$} & {\small  $5.67\times10^{-4}$} \\
\hline
{\small 1/year}  & {\small 11.35}   &  {\small 0.022}   & {\small 0.618}   
     & {\small 0.143} \\
\hline
\end{tabular}}
\end{center}
\vspace*{-1em}
\caption{{\small \label{paramVal}Parameters of the model obtained from the fit
  of the Dow-Jones data.  We also find $\rho=0$ for the correlation
  coefficient and $1/\gamma=22.2$ trading days for the relaxation time
  of variance.}}
\end{table}

The effective growth rate of stock prices is determined by the
coordinate $r_m(t)$ where the probability density $P_t(r_m)$ is
maximal.  Using the relation $r_m=x_m+\mu t$ and Eq.\ (\ref{x_m}), we
find that the actual growth rate is
$\bar\mu=\mu-\gamma\theta/2\omega_0=13$\% per year.  This number
coincides with the average growth rate of the Dow-Jones index obtained
by a simple fit of the time series $\{S_\tau\}$ with an exponential
function of $\tau$.  The effective stock growth rate $\bar{\mu}$ is
comparable with the average stock volatility after one year
$\sigma=\sqrt{\theta}=14.7\%$.  Moreover, the parameter (\ref{width}),
which characterizes the width of the stationary distribution of
variance, is equal to $\chi=0.54$.  This means that the distribution
of variance is broad, and variance can easily fluctuate to a value
twice greater than the average value $\theta$.  Thus, even though the
average growth rate of stock index is positive, there is a substantial
probability $\int_{-\infty}^0 dr\,P_t(r)=17.7$\% to have negative
growth for $t=1$ year.

According to (\ref{q_*}), the asymmetry between the slopes of
exponential tails for positive and negative $x$ is given by the
parameter $p_0$, which is equal to 1/2 when $\rho=0$ (see also the
discussion of Eq.\ (\ref{Pfinal'}) at the end of Sec.\ \ref{t>>1}).  The
origin of this asymmetry can be traced back to the transformation from
(\ref{eqS}) to (\ref{eqR}) using It\^{o}'s formula.  It produces the
term $0.5v_t\,dt$ in the r.h.s.\ of (\ref{eqR}), which then generates
the second term in the r.h.s.\ of (\ref{FP}).  The latter term is the
only source of asymmetry in $x$ of $P_t(x)$ when $\rho=0$.  However,
in practice, the asymmetry of the slopes $p_0=1/2$ is quite small
(about 2.7\%) compared with the average slope
$q_*^\pm\approx\omega_0/\kappa=18.4$.

\subsection{Conclusions}

We derived an analytical solution for the probability distribution
$P_t(x)$ of log-returns $x$ as a function of time $t$ for the model of
a geometrical Brownian motion with stochastic variance.  The final
result has the form of a one-dimensional Fourier integral
(\ref{Pfinal}) and (\ref{phaseF}).  [In the case $\rho=0$, the
equations have the simpler form (\ref{Pfinal'}), (\ref{phaseF'}), and
(\ref{eqOmega'}).]  Numerical evaluation of the integral
(\ref{Pfinal}) is simple compared with computationally-intensive
numerical solution of the original Fokker-Planck PDE or Monte-Carlo
simulation of the stochastic processes.

Our result agrees very well with the Dow-Jones data, as shown in Fig.\
\ref{fig:data}.  Comparing the theory and the data, we determine the
four (non-zero) fitting parameters of the model, particularly the
variance relaxation time $1/\gamma=22.2$ days.  For time longer than
$1/\gamma$, our theory predicts scaling behavior (\ref{Pbess}) and
(\ref{K1arg}), which the Dow-Jones data indeed exhibits over seven
orders of magnitude, as shown in Fig.\ \ref{fig:Bessel}.  The scaling
function $P_{\ast}(z)=K_1(z)/z$ is expressed in terms of the
first-order modified Bessel function $K_1$.  Previous estimates in
literature of the relaxation time of volatility using various indirect
indicators range from 1.5 days \cite[p.\ 80]{Papanicolau} to less than
one day and more than few tens of days \cite[p.\ 70]{Bouchaud-book}
for S\&P 500 and 73 days for the half-life of the Dow-Jones index
\cite{Engle}.  Since we have a very good fit of the entire probability
distribution function for times from 1 to 250 trading days, we believe
that our estimate, 22.2 days, is much more reliable.  A close value of
19.6 days was found in Ref.\ \cite{Masoliver}.

As Fig.\ \ref{fig:data} shows, the probability distribution $P_t(x)$
is exponential in $x$ for large $|x|$, where it is characterized by
time-dependent slopes $d\ln P/dx$.  The theoretical analysis presented
in Sec.\ \ref{x>>1} shows that the slopes are determined by the
singularities of the function $F_t(p_x)$ from (\ref{phaseF}) in the
complex plane of $p_x$ that are closest to the real axis.  The
calculated time dependence of the slopes $d\ln P/dx$, shown in Fig.\
\ref{fig:slope}, agrees with the data very well, which further
supports our statement that $1/\gamma=22.2$ days.  Exponential tails
in the probability distribution of stock log-returns have been noticed
in literature before \cite[p.\ 61]{Bouchaud-book}, \cite{Miranda},
however time dependence of the slopes has not been recognized and
analyzed theoretically.  In Ref.\ \cite{Stanley-returns}, the
power-law behavior of the tails has been emphasized.  However, the
data for S\&P 500 were analyzed in Ref.\ \cite{Stanley-returns} only
for short time lags $t$, typically shorter than one day.  On the other
hand, our data analysis is performed for the time lags longer than one
day, so the results cannot be directly compared.

Our analytical expression for the probability distribution of returns
can be utilized to calculate option pricing.  Notice that it is not
necessary to introduce the ad-hoc function $\lambda(S,v,t)$, the
market-price of risk, as is typically done in literature
\cite{Wilmott,Papanicolau,Group,SteinStein,Heston}, because now
$P_t(x)$ is known explicitly.  Using the true probability distribution
of returns $P_t(x)$ should improve option pricing compared with the
previous efforts in literature.

Although we tested our model for the Dow-Jones index, there is nothing
specific in the model which indicates that it applies only to stock
market data.  It would be interesting to see how the model performs
when applied to other time series, for example, the foreign exchange
data \cite{Peinke}, which also seem to exhibit exponential tails.

\section{Path-integral solution of the Cox-Ingersoll-Ross/Feller model}

In this section we provide details on the path-integral solution of
the variance process (\ref{eqVar}).  The stochastic differential
equation of the Cox-Ingersoll-Ross/Feller process, Eq.\ (\ref{eqVar}),
is
\begin{equation}\label{stochCIR}
 dx_t = -\gamma(x_t - \theta)dt + \kappa\sqrt{x_t}dW_t. 
\end{equation}
Here $\theta$ is the long-time mean of $x$, $\gamma$ is the rate of
relaxation to this mean, $W_t$ is the standard Wiener process, and
$\kappa$ is a parameter that controls the amplitude of the noise.
Eq.\ (\ref{stochCIR}) is known in financial literature as the
Cox-Ingersoll-Ross (CIR) process and it was used to model of interest
rate dynamics.  Feller \cite{Feller2} used equation (\ref{stochCIR})
to model population extinction in biological systems.  Recently, the
same equation (\ref{stochCIR}) was used to model the thermal reversal
of magnetization in a magnetic grain \cite{Safonov}.

One quantity of interest to calculate is the conditional (transition)
probability $P(x,t\,|\,x_i,t_i)$, which gives the pro\-bability to find
the system at position $x$ at time $t$, given that it was at position
$x_i$ at initial time $t_i$. The transition probability
$P(x,t\,|\,x_i,t_i)$ of the stochastic process described by
(\ref{stochCIR}) satisfies the Fokker-Planck equation
\begin{eqnarray} \label{FPApp}
  && \frac{\partial}{\partial t}P(x,t\,|\,x_i,t_i)
   =\frac{\partial}{\partial x}\left[\gamma(x-\theta)
   P(x,t\,|\,x_i,t_i)\right]  \nonumber \\
  && {}+\frac{\kappa^2}{2}\frac{\partial^2}{\partial^2 x}
    \left[xP(x,t\,|\,x_i,t_i)\right]
\end{eqnarray}
The Fokker-Planck equation (\ref{FPApp}) can be though of as a
Schr\"{o}dinger equation in imaginary (Euclidean) time written in the
coordinate ``x'' representation,
\begin{equation}
   \frac{\partial}{\partial t}P(x,t\,|\,x_i,t_i) = 
   - \hat{H}(\hat{p},\hat{x})P(x,t\,|\,x_i,t_i)
\end{equation}
with the Hamiltonian
\begin{equation} \label{HApp}
   \hat{H}=
   \frac{\kappa^2}{2}\hat{p}^2\hat{x}-i\gamma\hat{p}(\hat{x}-\theta). 
\end{equation}
In the coordinate representation, the momentum operator $\hat{p}$ is
given by the derivative operator $\hat{p}=-id/dx$. The canonical
commutation relation holds true: $[\hat{x},\hat{p}]=i$. Using this
analogy with quantum mechanics, the transition pro\-ba\-bi\-li\-ty
$P(x,t\,|\,x_i,t_i)$ can be found using the method of path integrals.
The transition probability is the matrix element of the evolution
operator $\hat{U}(t,t_i)=\exp(-\hat{H}(t-t_i))$ between the initial
state $|\,x_i\rangle$, and final state $\langle x_f\,|$,
\begin{equation}
  P(x_f,t_f\,|\,x_i,t_i)=\langle x_f\,|\,e^{-\hat{H}(t_f-t_i)}
  \,|\,x_i\rangle. 
\end{equation}

The standard construction of the path integral is obtained by 
splitting the time interval $[t_i,t_f]$ into M time steps of 
equal size $\epsilon=(t_f-t_i)/M$. 
One uses the multiplication property of the transition probability to write 
\begin{eqnarray} \label{infPropVV}
  && \langle x_f|e^{-\hat{H}(t_f-t_i)}|x_i\rangle=\int
     \prod_{j=1}^{M-1}(dx_j) \nonumber \\
  && \times\prod_{j=1}^{M}\langle x_{j}|e^{-\hat{H}(t_{j}-t_{j-1})}
     |x_{j-1}\rangle.
\end{eqnarray} 
The contribution coming from interval $(t_j,t_{j+1})$ is evaluated as
\begin{eqnarray} \label{infPropPV}
  && \langle x_j|e^{-\hat{H}(t_j-t_{j-1})}|x_{j-1}\rangle  = 
   \int \!\!dp_j \langle x_j|p_j\rangle \nonumber \\  
  &&\times \langle p_{j}|e^{-\hat{H}(t_{j}-t_{j-1})}|x_{j-1}\rangle.
\end{eqnarray}
The matrix element inside the integral in Eq.(\ref{infPropPV}) can be now 
evaluated to 
\begin{equation} \label{pvPart}
   \langle p_{j}|e^{-\hat{H}(t_{j}-t_{j-1})}|x_{j-1}\rangle \approx  
   e^{-\epsilon H(p_j,x_{j-1})}\langle p_{j}|x_{j-1}\rangle.
\end{equation}
Substitute the scalar product  
\begin{equation} \label{scalProd}
   \langle x_j|p_j\rangle = \frac{e^{ip_jx_j}}{\sqrt{2\pi}}
   \quad\quad\quad 
   \langle p_j|x_{j-1}\rangle = \frac{e^{-ip_jx_{j-1}}}{\sqrt{2\pi}},
\end{equation}
and Eq.(\ref{pvPart}) back into (\ref{infPropPV}), to obtain
\begin{eqnarray} \label{discretePI}
   && P(x_f,t_f\,|\,x_i,t_i)=\int\prod_{j=1}^{M-1}dx_j
      \int\prod_{j=1}^{M}\frac{dp_j}{2\pi}\,  \nonumber \\
   && \times e^{\sum_{j=1}^{M} 
      ip_j(x_j - x_{j-1}) - \epsilon H(p_j,x_{j-1})}.
\end{eqnarray}
where 
\begin{equation}
  H(p_j,x_{j-1})=\frac{\kappa^2}{2}p_j^2x_{j-1}-i\gamma p_j(x_{j-1}-\theta)
\end{equation}
By taking the limit $M\rightarrow\infty$ in Eq.(\ref{discretePI}), one
arrives at the path integral expression
\begin{equation}\label{contPI}
  P(x_f,t_f\,|\,x_i,t_i)= 
  \int\!\! {\cal{D}}x\,{\cal{D}}p\,e^{S[x,p]}
\end{equation}
where the action functional $S[x_t,p_t]$ is given by      
\begin{equation} \label{actionApp}
   S[x_t,p_t]=\int_{t_i}^{t_f}\!\! dt\left[ 
   ip_t\dot{x}_t-\frac{\kappa^2}{2} p_t^2 x_t
   +i\gamma p_t\left(x_t-\theta\right)\right].   
\end{equation}
The phase-space path integral from Eq.(\ref{contPI}) has to be calculated 
over all continuous paths $p(t)$ and $x(t)$ satisfying the
boundary conditions $x(t_i)=x_i$ and $x(t_f)=x_f$.

For most path-integrals in physics, the path integral is quadratic in
momenta and because of that it is customary to integrate first the
momenta $p(t)$ to obtain the Lagrangean of the problem, and be left
with the path integral over the coordinate $x(t)$. This route can be
taken here too, but the resulting Lagrangean will have a position
dependent mass, which unnecessary complicates the remaining
integration over the coordinate. It is important to notice that the
problem is linear in the coordinate $x(t)$ and quadratic in $p(t)$, so
the original Schr\"{o}dinger formulation will be simpler in the
momentum representation.  In the path-integral formulation, it turns
out that the integration over $x(t)$ is trivial and gives a delta
functional.

Both the continuous expression (\ref{contPI}) and the discrete one
(\ref{discretePI}) have their own computational advantages. In the
following, I will use the discrete form (\ref{discretePI}), because
for this problem, it allows for a quick and transparent solution. 
The exponent in (\ref{discretePI}) can be rearranged in the form 
\begin{eqnarray}
   && S=i(p_M x_f - p_0 x_i) - i\epsilon\gamma\theta
      \sum_{j=1}^{M}p_j  \\ \nonumber 
   && {}+i\sum_{j=1}^{M-1}x_j\left[(p_{j-1}-p_j) + \epsilon\gamma p_j  
   +i\epsilon\frac{\kappa^2}{2}p_j^2\right]
\end{eqnarray}
where I made the notation 
$p_0=p_1-\epsilon(\gamma p_1 +i\kappa^2p_1^2/2)$. 

The path integral over the coordinate can be taken one by one,
starting with the one over $dx_{M-1}$.  The result is a delta function
$2\pi\delta(p_{M-1}-p_M+\epsilon\gamma p_{M} + i\epsilon\kappa^2
p_M^2/2)$.  The integration over $p_{M-1}$ resolves the delta function
for the value of $p_{M-1}$ equal to $p_{M-1}\equiv(1 - \epsilon
\hat{T})p_M$, where the operator $\hat{T}=\gamma(\cdot) + k^2
(\cdot)^2/2$. The integral over $dx_{M-2}$ is done next, with the
result $2\pi\delta(p_{M-2}-(1-\hat{T})^2p_M)$, followed by the
integral over $dp_{M-2}$. The procedure is repeated until the last
pair $dx_1\,dp_1$. In this way we get $p_0=(1-\epsilon\hat{T})p_1=
(1-\epsilon\hat{T})^Mp_M$.

The value of the transition probability is 
\begin{equation} \label{finalDiscrete}
   P(x_f,t_f\,|\,x_i,t_i) = \int_{-\infty}^{\infty}\frac{dp_M}{2\pi} 
   e^{i(p_M x_f - p_0 x_i) - i\epsilon\gamma\theta\sum_{j=1}^{M}p_j}
\end{equation}
where in (\ref{finalDiscrete}),
$p_j\equiv(1-\epsilon\hat{T})^{(M-j)}p_M$ for
$j=\overline{0,M}$. Because of the approximation used in
(\ref{pvPart}), Eq.(\ref{finalDiscrete}) is valid only when the number
of time splittings $M\to\infty$. In this limit we have
\begin{equation} \label{discLim}
   \frac{p_j-p_{j-1}}{\epsilon}=\hat{T}p_j \quad\longrightarrow\quad 
   \frac{dp_t}{dt}=\hat{T}p_t\quad\mbox{with}\quad p(t_f)=p_M
\end{equation}
The differential equation $dp_t/dt=\hat{T}p_t=\gamma p_t
+i\frac{\kappa^2}{2}p_t^2$ is a Bernoulli equation and has the
solution
\begin{equation} \label{solBern}
   p_t=\frac{1}{(\frac{1}{p_M}+\frac{i\kappa^2}{2\gamma})e^{-\gamma(t-t_f)}
   -\frac{i\kappa^2}{2\gamma}}.
\end{equation}
The \hfill expression \hfill for \hfill the \hfill transition \hfill 
probability \hfill in \hfill the \\ $M\to\infty$ limit is 
\begin{equation} \label{finalCont}
 P(x_f,t_f\,|\,x_i,t_i) = \int_{-\infty}^{\infty}\frac{dp_M}{2\pi} 
   e^{i(p_M x_f - p_0 x_i) - i\gamma\theta\int_{t_i}^{t_f}dt\,p_t}
\end{equation}
where $p_t$ is given by Eq.(\ref{solBern}). The time integral in 
(\ref{finalCont}) can be evaluated as 
\begin{equation}
   \int_{t_i}^{t_f}dt\,p_t = 
   -\frac{2i}{\kappa^2}\int_0^{\gamma T}\frac{dy}{\zeta e^y -1}=
   -\frac{2i}{\kappa^2}\ln\frac{\zeta - e^{-\gamma T}}{\zeta-1}
\end{equation}
where
\begin{equation}
  \zeta=1-\frac{2i\gamma}{\kappa^2 p_M}, \quad\mbox{and}
  \quad T=t_f-t_i.
\end{equation}
After minor rearrangements in the argument of the logarithm, the final 
result is given in the form of a Fourier integral 
\begin{eqnarray} \label{finalFour}
  && \!\!\!\!\!\!\!\!\!\!\!\! P(x_f,t_f\,|\,x_i,t_i) = 
     \int_{-\infty}^{\infty}\frac{dp_M}{2\pi}
     e^{i(p_M x_f - p_0 x_i)} \nonumber \\ 
  && \!\!\!\!\!\!\!\!\!\!\!\! 
     \times \exp\left\{- \frac{2\gamma\theta}{\kappa^2}
     \ln\left[1 + \frac{i\kappa^2}{2\gamma}(1-e^{-\gamma T})p_M\right]
     \right\}
\end{eqnarray}
where 
\begin{equation}
   p_0\equiv p(t=t_i)=
   \frac{p_Me^{-\gamma T}}{1+\frac{i\kappa^2}{2\gamma}(1-e^{-\gamma T})p_M}
\end{equation}

\begin{figure}
\centerline{
\epsfig{file=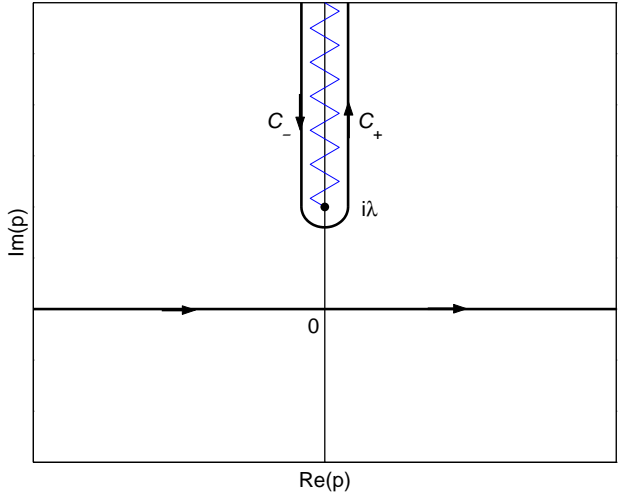,width=0.95\linewidth}}
\caption{{\small The complex plane of $p$. The branch cut due to the 
   denominator of Eq.\ (\ref{Four2}) is shown as a broken line.  
   The contour $\mathcal{C}$ has been deformed so that it encircles 
   the branch cut.}}
\label{fig:complexApp}
\end{figure}

Although, the integral from Eq.(\ref{finalFour}) seems complicated,
it can be evaluated exactly by integration in the complex
plane. Introduce the following notations 
\begin{equation} \label{param}
   \nu=2\gamma\theta/\kappa^2, \quad\mbox{and}\quad 
   \lambda=2\gamma/\kappa^2(1-\exp(-\gamma T)). 
\end{equation}
The integral in (\ref{finalFour}) becomes 
\begin{equation} \label{Four2}
   P(x_f,t_f\,|\,x_i,t_i) = \int_{-\infty}^{+\infty}\frac{dp}{2\pi}\, 
   \frac{\exp\left\{ipx_f - i\frac{pe^{-\gamma T}}{1+ip/\lambda}x_i\right\}}
   {(1 + ip/\lambda)^\nu}
\end{equation}
There is a branch point singularity in the complex plane of $p$ which
is situated on the imaginary axis, for $p=i\lambda$.  We take the
branch cut to extend from $i\lambda$ to $i\infty$.  For positive $x_f$,
the contour of integration is deformed so that it encircles the branch
cut.  The complex plane of $p$, and the contour of integration are
presented in FIG.\ref{fig:complexApp}.  The contribution from the
contour $\mathcal{C}$ can be split into two parts, one from the part
to the left of the cut $\mathcal{C}_-$ and the other part, to the
right of the cut $\mathcal{C}_+$.
The Bessel function $J_{\nu}(z)$ has the following integral
representation due to Schl\"afli
\begin{equation}
   J_{\nu}(z)=\frac{1}{2\pi i}\left(\frac{z}{2}\right)^{\nu}
   \int_{-\infty}^{(0+)}\!\! dt\, t^{-\nu-1}e^{t - z^2/4t},
\end{equation}
where the integral is taken along a contour around the branch cut on
the negative real axis, encircling it in the counterclockwise
direction. By making the change of variable \hfill $p\to -ip$ \hfill 
in \hfill (\ref{Four2}), \hfill the \hfill transition \hfill 
probability \\ $ P(x_f,t_f\,|\,x_i,t_i)$ can be expressed
in terms of the Bessel function as 
\begin{eqnarray}
  && P(x_f,t_f\,|\,x_i,t_i)=\lambda^\nu e^{-\lambda(x_f+x_ie^{-\gamma T})}
                                         \\  \nonumber
  && \times \left(-\frac{x_i}{x_f}\lambda^2 
     e^{-\gamma T}\right)^{\frac{1-\nu}{2}}  
     J_{\nu-1}(2i\sqrt{x_ix_f\lambda^2\exp(-\gamma T)}).
\end{eqnarray}
The modified Bessel function $I_{\nu}(z)$ is defined in terms of the
usual Bessel function $J_{\nu}(z)$ of imaginary argument by the
relation
\begin{equation}
   I_{\nu}(z) = e^{-\frac{i\pi\nu}{2}}J_{\nu}(e^{i\pi/2}z)\quad
   \mbox{for}\quad -\pi< arg(z) \le \pi/2. 
\end{equation}
This relation allows to write \textbf{the final result} as
\begin{eqnarray} \label{final}
  && P(x_f,t_f\,|\,x_i,t_i)=\lambda^\nu e^{-\lambda(x_f+x_ie^{-\gamma T})}
     \\  \nonumber
  && \times \left(\frac{x_i}{x_f}\lambda^2 
     e^{-\gamma T}\right)^{\frac{1-\nu}{2}}
     I_{\nu-1}(2\sqrt{x_ix_f\lambda^2\exp(-\gamma T)}), 
\end{eqnarray}
were $\lambda$ and $\nu$ are given by (\ref{param}), and $T=t_f-t_i$.

The normalization of the probability density, 
$$\int_0^\infty dx_f\,P(x_f,t_f\,|\,x_i,t_i) = 1,$$ 
can be proved easily by making use of the result
\begin{equation}
   \int_0^\infty \!\!dx\, x^{\nu+1} e^{-\alpha x^2} I_{\nu}(\beta x)=
   \frac{\beta^\nu}{(2\alpha)^{\nu+1}}e^{\beta^2/4\alpha}.
\end{equation}

We can check the validity of expression (\ref{final}) by taking
several limiting cases:

$\bullet$ In \hfill the \hfill limit \hfill $T\to\infty$, \hfill 
the \hfill transition \hfill probability \\
$P(x_f,t_f\,|\,x_i,t_i)$ should become the stationary probability
distribution (given in our paper).  The argument of the Bessel
function from (\ref{final}) goes to zero in the limit $T\to\infty$. 
The modified Bessel function $I_\nu(z)$ has the following small
argument behavior
\begin{equation} \label{smallArg}
   I_{\mu}(z)\approx\frac{1}{\Gamma(\mu+1)}
   \left(\frac{z}{2}\right)^{\mu} \quad\mbox{for}\quad z\to 0^+.
\end{equation}
Substituting the expression (\ref{smallArg}) into (\ref{final}), we
obtain
\begin{equation} \label{sinGamma}
   P(x_f,t_f\,|\,x_i,-\infty)\equiv P_{\ast}(x_f)=
   \frac{\lambda^\nu}{\Gamma(\nu)}x_f^{\nu-1}e^{-\lambda x_f}.
\end{equation}
which is the stationary distribution. 

$\bullet$ In the short time limit, $T\to 0$ $(\lambda\to\infty)$, the
probability distribution should approach a delta function,
$P(x_f,t_i\,|\,x_i,t_i)=\delta(x_f-x_i)$. The argument of the Bessel
function from (\ref{final}) goes to infinity in the limit $T\to
0$. The modified Bessel function $I_\nu(z)$ has the following large
argument behavior
\begin{equation} \label{largeArg}
   I_{\mu}(z)\approx \frac{e^{z}}{\sqrt{2\pi z}}
   \quad\mbox{for}\quad z\to\infty
\end{equation}
Substituting (\ref{largeArg}) back into (\ref{final}), we obtain
\begin{eqnarray}
   P(x_f,t_f\,|\,x_i,t_i)\!&\!=\!&\!f(x_i/x_f)
   \lim_{\lambda\to\infty}
   \sqrt{\frac{\lambda}{4\pi x_i}}\,
   e^{-\lambda(\sqrt{x_f}-\sqrt{x_i})^2} \nonumber \\
   \!&\!=\!&\! \delta(x_f-x_i)
\end{eqnarray}




\begin{thebibliography}{99}
\bibitem{Tsallis} It is also implied that the system is extensive.  We
   study only extensive models, so the nonextensive generalization of
   entropy by C. Tsallis, J. Stat. Phys. {\bf 52}, 479 (1988);
   Braz. J. Phys. {\bf 29}, 1 (1999) is not relevant for our
   consideration.
\end{thebibliography}
\end{document}